\documentclass[aps,pre, twocolumn,showpacs]{revtex4}
\usepackage{amsmath}
\usepackage{epsfig}
\usepackage{amssymb}
\usepackage{dcolumn}
\usepackage{graphicx}
\usepackage{dcolumn}

\begin{document}
\date{\today}

\title{Defect modes of a Bose-Einstein condensate in an optical lattice with a localized impurity}

\author{Valeriy A. Brazhnyi} 
\email{brazhnyi@cii.fc.ul.pt}
\affiliation{Centro de F\'{\i}sica Te\'{o}rica e Computacional, Faculdade de Ci\^encias, Universidade de Lisboa, 
Complexo Interdisciplinar, Avenida Professor Gama Pinto 2, Lisbon 1649-003, Portugal}
  
\author{Vladimir V. Konotop}
\email{konotop@cii.fc.ul.pt}
\affiliation{Centro de F\'{\i}sica Te\'{o}rica e Computacional, Faculdade de Ci\^encias, Universidade de Lisboa, 
Complexo Interdisciplinar, Avenida Professor Gama Pinto 2, Lisbon 1649-003, Portugal and Departamento de F\'{\i}sica, Faculdade de Ci\^encias,
Universidade de Lisboa, Campo Grande, Edif\'{\i}cio C8, Piso 6, Lisboa
1749-016, Portugal
}
 
\author{V\'{\i}ctor M. P\'erez-Garc\'{\i}a}
\email{victor.perezgarcia@uclm.es}
\affiliation{Departamento de Matem\'aticas, Escuela T\'ecnica
Superior de Ingenieros Industriales, 
Universidad de Castilla-La Mancha 13071 Ciudad Real, Spain}


\begin{abstract}

We study defect modes of a Bose-Einstein condensate in an optical lattice with a localized defect within the framework of the one-dimensional Gross-Pitaevskii equation. 
It is shown that for a significant range of parameters the defect modes can be accurately described by an expansion over Wannier functions, whose envelope is governed by the coupled nonlinear Schr\"{o}dinger equation with a delta-impurity. The stability of the defect modes is verified by direct numerical simulations of the underlying Gross-Pitaevskii equation with a periodic plus defect potentials. We also discuss possibilities of driving defect modes through the lattice and suggest ideas for their experimental generation.
\end{abstract}


\maketitle

\section{Introduction}

The recent progresses of experimental studies of Bose-Einstein condensates (BECs) in optical lattices (OLs)~\cite{Morsch} and in particular, the direct observation of a gap soliton~\cite{GapSol} have stimulated many studies of the dynamics of matter waves in periodic media (see e.g. the reviews \cite{Morsch,review} and references therein).  In relation with nonlinear matter waves, OLs provide a tool for changing the effective properties of the atomic medium allowing the existence of solitary waves in condensates with positive scattering length. Moreover, deformed OLs offer additional possibilities for manipulating the soliton dynamics. In particular, smoothly modulated lattices are effective in controlling the quasi-one-dimensional dynamics of  small amplitude gap solitons~\cite{BKK,review}.

The study of various types of localized defects and their interaction with matter waves has been considered in different contexts, ranging from experimental studies~\cite{10} and different theoretical works studying the emission of vortices in the superfluid flow of a BEC around a defect~\cite{9}, the generation of gray solitons and sound waves by a moving defect~\cite{11}, the drag force appearing at large velocities of a moving defect in a BEC~\cite{AstPit}, and the combined effect of a static defect and trapping potential~\cite{14}.  

In Ref.~\cite{ours} we have explored theoretically some features of the stationary configurations of a BEC under the simultaneous effect of an OL and of a strongly localized defect. This leads to the existence of localized states, termed {\em nonlinear defect modes}, which can be driven through hundreds of periods of an OL. In this paper we extend the analysis of Ref. \cite{ours}, addressing the problems of existence, stability, and generation of stationary modes localized in the vicinity of the defect, and consider the possibility of the dynamical interactions between defect modes. We provide details of an approximate analytical description of defect modes, outlined in~\cite{ours} and report thorough numerical simulations of their dynamics in order to understand their stability properties.  To do so,  we will exploit the fact that in the presence of an OL, the wave function of the BEC can be expanded on a set of Wannier functions (WFs)~\cite{AKKS,KAKS,review} localized on each potential well \cite{kohn} and modulated by an envelope. 

Before going into details, we would like to mention that the theory of nonlinear defect modes in one-dimensional (1D) periodic media, is an issue of high relevance also for the theory of photonic crystals. The case of lattices with shallow grating was recently addressed in~\cite{Goodman1} within the framework of the coupled-mode equations for the counter-propagating waves. In the same approximation, light propagation in nonlinear fibers with slowly modulated Bragg gratings was considered in Ref.~\cite{Mak}. The research reported in the present paper, as well as in earlier publications~\cite{BKK,ours}, differs not only by the physical model, but also by the fact that the grating is considered to be deep and the dynamics is studied numerically within the full 1D equation. 

In this context it is also relevant to mention studies of impurity modes in the framework of the discrete nonlinear Schr\"odinger equation~\cite{dp1}, since in the tight-binding approximation, i.e. in the limit opposite to shallow lattices, the nonlinear Schr\"odinger (NLS) equation  with a periodic potential can be mapped into the discrete NLS equation~\cite{AKKS,KAKS,review}.  
 
The structure of this paper is as follows: In Sec.~\ref{II} we expand the wave function in terms of the WF's and derive coupled equations governing the evolution of the envelope. Then, in Sec. \ref{III} we provide the simplest analytical solutions corresponding to stationary defect modes. We also study the dynamics and stability of these modes by direct numerical simulations of the 1D Gross-Pitaevskii (GP) equation. Another interesting question is the possibility of moving these defect modes, which we consider in Sec.~\ref{IV}. In Sec.~\ref{V} we propose several ideas for the generation of defect modes in realistic experiments. Finally, in Sec.~\ref{VI} we summarize the outcomes.

\section{Statement of the problem and analytical consideration}
\label{II}
\subsection{Bloch and Wannier function expansions}

We consider a BEC trapped in a quasi-1D configuration, i.e. tightly confined  by a trap potential in the transverse direction, providing a cigar-shaped geometry. We also suppose that the system is subject to a 1D periodic potential, which is created by laser beams and whose axis is directed along the longitudinal direction of the trap. Provided the density of atoms is small, such a system is governed by the 1D GP equation (see e.g. ~\cite{Victor,review} for the details of derivation), which in dimensionless variables reads
\begin{equation}
\label{NLS}
i\frac{\partial\psi}{\partial t} = -\frac{\partial^2\psi}{\partial x^2}+V_l(x)\psi+V_d(x)\psi + \sigma |\psi|^{2}\psi.
\end{equation}
Here  $\sigma=1$ stands for repulsive and $\sigma=-1$ for attractive atom-atom interactions, $V_l(x)$ is an OL potential  and $V_d(x)$ is a defect potential. The dimensionless variables are chosen so that the energy is measured in the units of the recoil energy $E_r=\hbar^2k^2/(2m)$ where $k=\pi/d$ and $d$ is the lattice constant. Then the temporal and spatial coordinates are measured in the units of $E_r/\hbar$ and of $d/\pi$, respectively, and the period of an OL potential is $\pi$: $V_l(x+\pi)=V_l(x)$. The dimensionless macroscopic wave function is normalized as follows $\int  |\psi|^2dx=8\pi N k |a_s|$, where $a_s$ is the scattering length. 

In order to provide  a formal derivation of the approximate equations we will assume that the defect potential is localized in space on a distance smaller than one lattice period, what mathematically can be expressed as $\mid$~$\frac{1}{V_d}\frac{dV_d}{dx} \mid\ll O(\pi^{-1})$ (the numerical studies presented bellow show that this region can be significantly enlarged). 
For the sake of simplicity, we will restrict our analysis to symmetric OL potentials $V_{l}(x)=V_{l}(-x)$ and represent the defect potential in the form $V_{d}(x)=V_{+}(x)+V_{-}(x)$, where $V_{\pm}(x)=\pm V_{\pm}(-x)$  are symmetric, $V_+$, and antisymmetric, $V_-$, parts, respectively.

In the absence of nonlinearity, Eq.~(\ref{NLS}) describes the well-known problem of a quantum particle in a periodic potential which arises, for instance, in description of the interaction of an electron with a defect in a 1D crystal lattice. We will make use of this fact and use
approximation techniques which are well developed in solid state physics~\cite{solid}.
 
To this end we introduce an expansion of the macroscopic wave function
\begin{eqnarray}
	\label{expan1}
	\psi(x,t)=\sum_{\alpha=0}^{\infty}\int_{-1}^1 dk \varphi_{\alpha k}(x)
	f_\alpha (k,t)
\end{eqnarray}
over the orthonormal set of the Bloch functions $\varphi_{\alpha k}(x)$ of the linear eigenvalue problem
\begin{eqnarray}
\label{eigen}
	-\frac{d^2\varphi_{\alpha k}}{dx^2}+V_l(x)\varphi_{\alpha k}=E_\alpha (k)\varphi_{\alpha k}.
\end{eqnarray}
Hereafter $\alpha$ stands for the number of the zone, subject to convention that the lowest allowed band has zero number, $k$ is the wavevector considered in the first Brillouin zone (BZ), $k\in[-1,1]$. Then it is a straightforward algebra to obtain from Eqs. (\ref{NLS})-(\ref{eigen}) the following equation for the envelope $f_\alpha (k,t)$:
\begin{eqnarray}
	\label{eq1}
	&& i\dot{f}_\alpha (k,t)-E_\alpha(k)f_\alpha (k,t) 
	=\sigma\int dx 
	\bar\varphi_{\alpha k}(x)|\psi|^2\psi(x,t)
	\nonumber \\
	&&+\sum_{\alpha'}\int dk'
	\int dx\bar{\varphi}_{\alpha k}(x)V_d(x)\varphi_{\alpha' k'}(x)
	f_{\alpha'} (k',t). 
\end{eqnarray}
In this paper we drop the limits in all integrals, taking the convention that integrals with respect to $x$ 
are considered over the whole line and integrals 
with respect to $k$ are considered over the first BZ, i.e. 
$\int dx...\equiv \int_{-\infty}^{\infty} dx...$ and $\int dk...\equiv \int_{-1}^{1} dk...$.

Now we introduce WFs  
\begin{eqnarray}
\label{wf}
w_{\alpha n}(x)=\frac{1}{\sqrt{2}}\int dk \, \varphi_{\alpha k}(x)e^{-i\pi nk},
\end{eqnarray}
which make up a complete orthonormal basis and can be chosen real and exponentially
decaying~\cite{kohn}. 
Each WF is centered at $x=n\pi$, where $n$ is an integer. 
The Bloch functions can be obtained from the WFs from the inversion of Eq. (\ref{wf}) through the relation
\begin{eqnarray}
	\label{bf}
	\varphi_{\alpha k}(x)=\frac{1}{\sqrt{2}}\sum_{n=-\infty}^{\infty}w_{\alpha n}(x)e^{i\pi nk}.
\end{eqnarray}

Introducing the notation
\begin{eqnarray}
	\label{eq2}
	V_{\alpha\alpha'}^{nn'}=\frac{1}{2}\int dx V_d(x)w_{\alpha n}(x)w_{\alpha' n'}(x)  
\end{eqnarray}
and defining the  Fourier transform of $V_{\alpha\alpha'}^{nn'}$:
\begin{eqnarray}
\label{fourierV}
	\hat{V}_{\alpha\alpha'}^{kk'}=\sum_{n,n'}V_{\alpha\alpha'}^{nn'}e^{i\pi (k'n'-kn)}\,,
\end{eqnarray}
which possesses the evident symmetry $\hat{V}_{\alpha\alpha'}^{kk'}=\overline{\hat{V}_{\alpha'\alpha}^{k'k}}$, we rewrite the defect term in Eq. \eqref{eq1} in the form   
\begin{multline}
\label{eq3}
	\sum_{\alpha'}\int dk'
	\int dx\bar{\varphi}_{\alpha k}(x)V_d(x)\varphi_{\alpha' k'}(x)
	f_{\alpha'} (k',t) 
		\\
	=\sum_{\alpha^\prime}\int dk'\hat{V}_{\alpha\alpha'}^{kk'}f_{\alpha'}(k',t).
\end{multline}
In Eqs. (\ref{fourierV}) and (\ref{eq3}) we have used the convention that all sums with respect to $n$ and $\alpha$ are considered over all integers and all nonnegative integers, respectively, i.e. $\sum_{n}...\equiv\sum_{n=-\infty}^{\infty}...$ and $\sum_{\alpha}...\equiv\sum_{\alpha=0}^{\infty}...$.

Next we rewrite the nonlinear term in Eq.~(\ref{eq1})  as 
\begin{widetext}
\begin{multline}
	\frac{1}{4} \sum_{nn_1n_2n_3}\sum_{\alpha_1\alpha_2\alpha_3} 
	\int dk_1dk_2dk_3W_{\alpha\alpha_1\alpha_2\alpha_3}^{nn_1n_2n_3}
 e^{i\pi (k_2n_2+k_3n_3-kn-k_1n_1)}\overline{f}_{\alpha_1}(k_1,t)
	f_{\alpha_2}(k_2,t) f_{\alpha_3}(k_3,t) = \\ 	\label{nl1}
	\frac{1}{2}\sum_{n_1n_2n_3}\sum_{\alpha_1\alpha_2\alpha_3} 
	\int_{BZ}dk_1dk_2dk_3\delta_{k+k_1,k_2+k_3} W_{\alpha\alpha_1\alpha_2\alpha_3}^{0n_1n_2n_3}
	 e^{i\pi (k_2n_2+k_3n_3-k_1n_1 )}\overline{f}_{\alpha_1}(k_1,t)
	f_{\alpha_2}(k_2,t) f_{\alpha_3}(k_3,t),
\end{multline}
\end{widetext}
where
\begin{eqnarray}
	\label{w}
	W_{\alpha\alpha_1\alpha_2\alpha_3}^{nn_1n_2n_3}=\int dx w_{\alpha n}w_{\alpha_1 
	n_1}w_{\alpha_2 n_2}w_{\alpha_3 n_3}.
\end{eqnarray}
In writing Eq. (\ref{nl1}) we have used the property that $W_{\alpha\alpha_1\alpha_2\alpha_3}^{nn_1n_2n_3}=W_{\alpha\alpha_1\alpha_2\alpha_3}^{0,n_1-n,n_2-n,n_3-n}$ (see \cite{AKKS,KAKS} for more details) and have computed the sum over $n$.
Now that we have rewritten Eq. (\ref{NLS}) in an appropriate form for our analysis we will discuss the approximate analytical description to be 
constructed in this paper.

\subsection{Constrains and approximations}

We will be interested in matter-waves of small amplitude, which are well described in the effective mass approximation~\cite{review,KS}, and thus having spatial extension, $\lambda$, much bigger than the lattice constant: $\lambda\gg \pi$. Then the respective characteristic scale $\Delta k\sim 2\pi/\lambda$ of $f_\alpha(k,t)$ in the momentum space is much smaller than the vector of the reciprocal lattice (since $\Delta k\cdot\lambda\sim 1$)
\begin{eqnarray}
	\label{cond}
	\Delta k\ll k_{BZ}=1\,.
\end{eqnarray}
Meantime the scale of the variation of $\hat{V}_{\alpha\alpha'}^{kk'}$ with respect to either of the arguments exceeds one.
In other words, we assume that $f_\alpha(k,t)$ is a strongly localized  function, compared with $\hat{V}_{\alpha\alpha'}^{kk'}$, the latter being weakly dependent on both $k$ and $k'$. Assuming also that $f_\alpha(k,t)$  is localized about some point $k_0$ in the first BZ (i.e. $|k_0|\leq k_{BZ}$) we can approximate 
\begin{eqnarray}
\label{eq4}
\int \hat{V}_{\alpha\alpha'}^{kk'}f_{\alpha'}(k',t)dk'\approx \hat{V}_{\alpha\alpha'}^{kk_0}\int f_{\alpha'}(k',t)dk'
\nonumber \\
\approx \hat{V}_{\alpha\alpha'}^{k_0k_0}\int f_{\alpha'}(k',t)dk'.
\end{eqnarray}

WFs are known to be exponentially localized in space~\cite{kohn} and 
the ``radius'' of localization decreases as the potential depth is increased, assuming the period remains  constant (for examples see e.g.~\cite{KAKS,review}).  

Subject to these conditions and taking into account the inequality (\ref{cond}) and the fact that $f(k,t)$ is localized about $k_0$ we conclude that the exponential factors $e^{in_jk_j\pi}$ in (\ref{nl1}) can be substituted by $e^{in_jk_0\pi}$ (with exponential accuracy), what allows us to introduce
\begin{eqnarray}
	\label{w1}
	W_{\alpha\alpha_1\alpha_2\alpha_3}=\frac{1}{2}\sum_{n_1n_2n_3} 
	W_{\alpha\alpha_1\alpha_2\alpha_3}^{0n_1n_2n_3}e^{i(n_2+n_3-n_1)k_0\pi}
\end{eqnarray}
and rewrite Eq. (\ref{nl1}) in the form
\begin{eqnarray}
	\label{nl11}
	 \sum_{\alpha_1\alpha_2\alpha_3} W_{\alpha\alpha_1\alpha_2\alpha_3}
	 \langle \overline{f}_{\alpha_1}(k_1,t)
	f_{\alpha_2}(k_2,t) f_{\alpha_3}(k_3,t)\rangle,
\end{eqnarray}
where, for the sake of brevity, we have introduced the notation
\begin{eqnarray}
	\label{average}
		\langle a(k_1)
	b(k_2) c(k_3)\rangle
	\equiv 
	\int_{BZ}dk_1 dk_2 dk_3 \delta_{k+k_1, k_2+k_3}  
	\nonumber \\
	\times
	a(k_1) 	b(k_2) c(k_3).
\end{eqnarray}

Whenever $V_{\alpha\alpha'}^{kk_0}\neq 0$ the defect term (\ref{eq4}) results in tunneling of atoms between the bands $\alpha$ and $\alpha'$.
The nonlinearity (\ref{w1}) is another  cause for inter-band tunneling. Therefore one can expect that a one-band approximation, successfully reproducing gap solitons in some homogeneous lattices \cite{AKKS}, may not work properly when the combined effect of the defect and the nonlinearity are considered. Below we will see, however, that this is not always so. 
 
We will also require that the potential $V_l(x)$ is such that the lowest even (including zero) and odd bands to have even and odd WFs, what, in particular, means that 
\begin{eqnarray}
\label{symW}
w_{\alpha n}(x)=(-1)^\alpha w_{\alpha n}(-x)\,.
\end{eqnarray}
This is the case of a symmetric OL potential, $V_l(x)=V_l(-x)$ having a local minimum at $x=0$: $V_l(x)\geq V_l(0)$. The imposed constrain is not strong, and in particular is satisfied by the cos-like potential [see (\ref{PerPot}) below] modeling most of the presently available experimental settings. 

Under the imposed conditions, one immediately obtains from (\ref{w}) that $W_{\alpha\alpha_1\alpha_2\alpha_3}^{nnnn}=0$ whenever $\alpha+\alpha_1+\alpha_2+\alpha_3=2p+1$ with $p$ being an integer. This means that the lattice originates an effective coupling of the bands with the same parity (i.e. for example atom transfer between band $\alpha=0$ and $\alpha=2$ or between bands $\alpha=1$ and $\alpha=3$). This fact leads us to the next simplification of the problem, considering  the {\em two-band approximation} assuming that initially all atoms possess energies in the two lowest bands, and thus only the bands (with $\alpha=0$ and $\alpha=1$) will contribute to the BEC dynamics. Thus we will impose that $f_{\alpha}(k,t)\ne 0$ for $\alpha=0,1$ and $f_{\alpha}(k,t)\equiv 0$ for $\alpha\geq 2$. Since this assumption is not supported by any rigorous statement (and the defect itself can result in the coupling of the bands) below the respective results will be compared with direct numerical simulations.

Then, taking only $\alpha,\beta =0,1$, $\beta\neq\alpha$ we obtain the final expressions for the nonlinear term 
\begin{widetext}
\begin{equation}
	\label{nl11_1}
		W_{\alpha\alpha} \langle \overline{f}_{\alpha}(k_1,t)	f_{\alpha}(k_2,t) f_{\alpha}(k_3,t)\rangle
	 +W_{\alpha\beta} \left( 2\langle \overline{f}_{\beta}(k_1,t)	f_{\alpha}(k_2,t) f_{\beta}(k_3,t)\rangle
	+\langle \overline{f}_{\alpha}(k_1,t)	f_{\beta}(k_2,t) f_{\beta}(k_3,t)\rangle\right),  
\end{equation}
Now defining $\hat{V}_{\alpha}\equiv 2\pi\hat{V}_{\alpha\,\alpha}^{k_0k_0}$, $\hat{V}\equiv 2\pi\hat{V}_{0\,1}^{k_0k_0}=2\pi\hat{V}_{1\,0}^{k_0k_0}$  and
$W_{\alpha\beta}=W_{\alpha\alpha\beta\beta}$ with $W_{0111}=W_{0001}= 0$ (the last two approximations are made on the basis of strong localization of the WFs resulting in the estimate $|W_{\alpha\alpha_1\alpha_2\alpha_3}^{0n_1n_2n_3}|\ll |W_{\alpha\alpha_1\alpha_2\alpha_3}^{0\,0\,0\,0}|$ if at least one of $n_j\neq 0$, see the examples in Table~\ref{table1} below) we obtain the equations
\begin{subequations}	
\label{eqnew}
\begin{eqnarray}
	i\dot{f}_0 (k,t)-E_0(k)f_0 (k,t)-\frac{\hat{V}_0}{2\pi} \int f_{0} (k',t)dk'
	-\frac{\hat{V}}{2\pi}\int f_{1} (k',t)dk'	 
	-\sigma W_{00} \langle \overline{f}_{0}(k_1,t)f_{0}(k_2,t)f_{0}(k_3,t)\rangle\nonumber\\
		- \sigma W_{01}\left[2\langle\overline{f}_{1}(k_1,t)f_{0}(k_2,t)f_{1}(k_3,t)\rangle 
		+\langle \overline{f}_{0}(k_1,t)f_{1}(k_2,t)f_{1}(k_3,t)\rangle\right]=0\,,
		\\
	i\dot{f}_1 (k,t)-E_1(k)f_1 (k,t)-\frac{\hat{V}_1}{2\pi} \int f_{1} (k',t)dk'
	-\frac{\hat{V}}{2\pi} \int f_{0} (k',t)dk'	 
	-\sigma W_{11} \langle\overline{f}_{1}(k_1,t)f_{1}(k_2,t)f_{1}(k_3,t)\rangle
	\nonumber\\
	- \sigma W_{10}\left[2\langle\overline{f}_{0}(k_1,t)f_{1}(k_2,t)f_{0}(k_3,t)\rangle 
	+\langle\overline{f}_{1}(k_1,t)f_{0}(k_2,t)f_{0}(k_3,t)\rangle\right]=0\,.
\end{eqnarray}
\end{subequations}
\end{widetext}
 
In the vicinity of the maxima of the functions $f_\alpha(k,t)$, i.e. in the vicinity of $k=k_0$,  one can expand
\begin{eqnarray}
\label{en_expan}
E_\alpha(k)=E_\alpha+v_\alpha (k-k_0)+ \frac{(k-k_0)^2}{2M_\alpha},
\end{eqnarray}
where $E_\alpha=E_\alpha(k_0)$, $v_\alpha=\left.\frac{dE_\alpha(k)}{dk}\right|_{k=k_0}$ is the group velocity of the mode $\alpha$, and  $M_\alpha=  \left(\frac{d^2E_\alpha(k)}{dk^2}\right)_{k=k_0}^{-1}$ is its effective mass.

\subsection{Dynamical equations}

Let us now introduce the function 
\begin{eqnarray}
\label{fourier}
\hat{f}_\alpha(x,t)=\int  e^{i(k-k_0)x}f_\alpha(k,t)\, dk.
\end{eqnarray}
Multiplying Eqs. (\ref{eqnew}) by $\exp[i(k-k_0)x]$ and integrating it with respect to $k$ over the BZ, we obtain a system of coupled  nonlinear Schr\"odinger (CNLS) equations with delta impurities
\begin{widetext}
\begin{subequations}
	\label{NLS-del}
\begin{eqnarray}
	i\frac{\partial \hat{f}_0}{\partial t}+iv_0\frac{\partial \hat{f}_0}{\partial x}+\frac{1}{2M_0} \frac{\partial^2 \hat{f}_0}{\partial x^2}-
	E_0\hat{f}_0 -	\sigma W_{00} |\hat{f}_0|^2\hat{f}_0
		-\sigma W_{01}\left(2 |\hat{f}_1|^2\hat{f}_0+\hat{f}_1^2\bar{\hat{f}}_0\right)	= \hat{V}_{0}\delta(x)\hat{f}_0+\hat{V}\delta(x)\hat{f}_1,
\\
	i\frac{\partial \hat{f}_1}{\partial t}+iv_1\frac{\partial \hat{f}_1}{\partial x}+\frac{1}{2M_1} \frac{\partial^2 \hat{f}_1}{\partial x^2}-E_1\hat{f}_1- 	\sigma W_{11} |\hat{f}_1|^2\hat{f}_1-	
	\sigma W_{01} \left( 2 |\hat{f}_0|^2\hat{f}_1+ \hat{f}_0^2\bar{\hat{f}}_1\right)
	=\hat{V}\delta(x)\hat{f}_0+ \hat{V}_{1}\delta(x)\hat{f}_1  .
		\end{eqnarray}
	\end{subequations}
\end{widetext}

These equations must be consistent with the expansion used for their derivation which implies that the effective masses should be of order of one, $|M_\alpha|\sim 1$.
These conditions are satisfied in reasonably large ranges of parameters for the most interesting  potentials from the physical point of view, when the depth of an OL is of the order of a few recoil energies.

To recover the wavefunction $\psi(x,t)$ from the solutions of Eqs. \eqref{NLS-del} we make use of Eqs. (\ref{expan1}) and (\ref{bf})  together with the definition (\ref{fourier}) and obtain
\begin{eqnarray}
	\label{recover}
	\psi(x,t)=
	\frac{1}{\sqrt{2}}\sum_{\alpha}\sum_ne^{i\pi k_0 n}
	w_{\alpha n}(x)\hat{f}_\alpha (n\pi,t).
\end{eqnarray}

In what follows we concentrate on the modes bordering the gap, i.e. modes with $k_0=k_{BZ}$ and with $v_\alpha=0$. In that particular case and taking into account that we are working in the two-band approximation, Eq. (\ref{recover}) takes the simple form
\begin{eqnarray}
	\label{recover_fin}
	\psi(x,t)&=&\frac{1}{\sqrt{2}}\sum_n(-1)^n\nonumber \\
	&\times&\left[w_{0n}(x)\hat{f}_0 (n\pi,t) +	w_{1n}(x)\hat{f}_1 (n\pi,t) \right].
\end{eqnarray}
Now we also have that, $\hat{V}_\alpha$ and $\hat{V}$ are originated by the symmetric and antisymmetric components of the defect, i.e.
\begin{subequations}
\label{V_approx}
\begin{eqnarray}
		\hat{V}_\alpha  & = &  \pi\sum_{n,n'}(-1)^{n+n'}\int V_+(x)w_{\alpha n}(x)w_{\alpha n'}(x)dx,\label{V_approx_a}
	\\
	\hat{V} & = & \pi\sum_{n,n'} (-1)^{n+n'} \int V_-(x)w_{0 n}(x)w_{1 n'}(x)dx.\label{V_approx_b}
\end{eqnarray}
\end{subequations}
Indeed, using that $w_{\alpha n}(x)\equiv w_{\alpha}(x-\pi n)$, i.e. the fact that WF's depend on $(x-\pi n)$, as well as the symmetry of WFs (\ref{symW}), we get 
\begin{multline}
	\int V(x)[w_{\alpha n}(x)w_{\alpha' n'}(x)+w_{\alpha,- n}(x)w_{\alpha',-n'}(x)]dx
	\\
	=
	\int [V(x)+(-1)^{\alpha+\alpha'}V(-x)]w_{\alpha n}(x)w_{\alpha' n'}(x) dx
\end{multline}
from which Eqs.\eqref{V_approx} follow.

Eqs. (\ref{V_approx}) imply that even defects couple only modes of the same parity (and in the two-band approximation does not result in any coupling), while the odd defect couples bands with different parity (i.e. the two adjacent bands in the two-band approximation).

\subsection{Lattice potential and numerical test of the approximations}

During the derivation of Eqs. (\ref{NLS-del}) a number of approximations have been used and several constrains have been imposed. 
That is why before going into details of solutions we will test the reliability of the analytical approximation. 
For the numerical calculations we will use the periodic potential in the standard form 
\begin{eqnarray}
	\label{PerPot}
	V_l(x)=A\cos(2x)
\end{eqnarray}
where $A$ is an amplitude of the lattice potential (we recall that it is measured in the units of the  recoil energy). 
By varying the amplitude of the OL one can control the magnitude of the parameters in Eqs.  (\ref{NLS-del}).

The first of our approximations has been to assume strong localization of the WF's.
We have also used an effective mass approximation which describes the behavior of the wave function near the boundary of the BZ what corresponds to the case of a small amplitude wavepacket. Thus, our theory is expected to be applicable for small amplitude wavepackets in relatively large amplitude OLs.

We have calculated the coefficients in Eqs. (\ref{NLS-del}) for different amplitudes of the OL, which are presented in Table~\ref{table1}~\footnote{The numerical results presented in this table were obtained using the software offered by G. L. Alfimov. For more details of the numerical study of the WFs and the related nonlinear coefficient for the potential (\ref{PerPot}) see \cite{AKKS}.}.
\begin{table*}[ht]
\caption{Effective mass, energy, and overlapping coefficients at the boundary of the BZ and at the edges of the first gap for different amplitudes of the OL. In parenthesis we indicate the signs and numbers of identical hopping integrals contributing to the nonlinearity $W_{\alpha\alpha_1\alpha_2\alpha_3}$ defined in (\ref{w1}).}
\begin{ruledtabular}
Lowest band, $\alpha=0$
\\
		\begin{tabular}{llll|llllll}
A& $M_0$   & $E_0$& $W_{00}$ & $W^{0000}_{0000}(1+)$ & $W^{0011}_{0000}(6+)$ & $W^{0001}_{0000}(8-)$ & $W^{0002}_{0000}(8+)$ & $W^{0112}_{0000}(6+)$ & $W^{0012}_{0000}(12-)$\\
\hline
$-1$& $-0.163$ & 0.471   & 0.2505&  0.3753    & 0.0082   & $-0.0064$    &	0.0035    & $-0.0026$     & $-0.0011$    \\
$-3$& $-0.85$ & $-0.733$ & 0.2983 &   0.5479    & 0.0006   & $-0.0052$    &	0.0005    & $-0.0001$     & $-0.0000$    \\
$-5$& $-2.5$ &$-2.076$ & 0.333  &   0.6479    & 0.0001   & $-0.0022$   &	0.0001    & 0.0000    & $-0.0000$    
		\end{tabular}
		\\
		First band, $\alpha=1$		
\\
		\begin{tabular}{llll|llllll}
A& $M_1$   & $E_1$& $W_{11}$  & $W^{0000}_{1111}(1+)$ & $W^{0011}_{1111}(6+)$ & $W^{0001}_{1111}(8-)$ & $W^{0002}_{1111}(8+)$ & $W^{0112}_{1111}(6+)$ & $W^{0012}_{1111}(12-)$\\
\hline
$-1$& $0.1$ & $1.467$  & 0.2069 & $0.2404$  & $0.0358$ & $-0.0163$ &	$-0.0008$ & $-0.0148$ &  $0.0064$ \\
$-3$& $0.21$ & $2.166$ & 0.2081 & $0.3236$  & $0.0211$ & $-0.0036$ &	$0.0035$ & $-0.0079$ &  $0.0035$ \\
$-5$& $0.3$  & $2.5$   & 0.2016 & $0.3976$   & $0.0075$ & $0.0048$ &	$0.0025$ & $-0.0020$ & $0.0008$    
		\end{tabular}
		\\
		Cross terms
\\
		\begin{tabular}{ll|lllllllllll}
A&  $W_{01}$ & $W^{0000}_{0011}$ & $W^{0011}_{0011}$ & $W^{0001}_{0011}$ & $W^{0002}_{0011}$ & $W^{0012}_{0011}$ & $W^{0022}_{0011}$ & $W^{0100}_{0011}$ & $W^{0101}_{0011}$ & $W^{0102}_{0011}$ & $W^{0112}_{0011}$ & $W^{0122}_{0011}$\\
\hline
$-1$& 0.1306 &  $0.1815$& $0.0399$ & $-0.0031$ &$0.0027$& $0.0073$& $0.0026$ & $0.0074$&$-0.0063$& $-0.0017$ & $0.0017$ & $-0.0029$ \\
$-3$& 0.1091 & $0.2221$	& $0.0105$	& $0.0089$& 0.0049& $0.0026$& $0.0007$& $0.0019$ & $-0.0018$& $-0.0003$& $0.0004$ 	& $-0.0001$ \\
$-5$& 0.1224 & $0.2703$	& $0.0028$	& $0.0100$& $0.0027$& $0.0005$& $0.0001$ & $0.0004$	& $-0.0004$	& $0.0000$& $0.0001$& $-0.0000$    
		\end{tabular}
	\label{table1}
	\end{ruledtabular}
\end{table*} 
As one can see from the table, the nonlinear coefficients in Eqs. (\ref{NLS-del}) are of different orders of magnitude. Specifically, $W_{\alpha\beta}\ll W_{\alpha\alpha}$ if $\alpha\neq \beta$, what is a natural consequence of the symmetry of the WFs. In fact, this is a property of a large class of potentials rather than a feature of our specific model (\ref{PerPot}). This approximation allows us to simplify significantly the problem by neglecting the  cross-phase-modulation terms, i.e. terms with $W_{01}$ in Eqs. (\ref{NLS-del}).

>From  Table \ref{table1} we also see that in  most of  the cases the effective mass is indeed of order of one, as it was supposed in Sec. \ref{II}. The explicit forms of the WFs for the cases considered in the table can be found in Ref.~\cite{KAKS}, where it is shown numerically that the WFs are indeed well localized on several (in the case $A=-1;-3$) periods of the OL, or even on a single period when $A=-5$.

\section{Stationary impurity modes}
\label{III}

\subsection{One-band approximation. Bright and dark defect modes}
\label{OneBand}

An important effect introduced by a localized potential is the coupling between different bands. When the defect has a well-defined parity, the problem is further simplified, as it follows from \eqref{V_approx}. Concentrating, first, on a defect of even symmetry, where $V_d(x)\equiv V_+(x)$ we obtain that $\hat{V}=0$ and the system of equations \eqref{NLS-del} becomes decoupled transforming in ($\alpha=0,1$):
\begin{multline}
	\label{NLS-del-sinlge}
 	i\frac{\partial \hat{f}_\alpha}{\partial t}
 	+\frac{1}{2M_\alpha} \frac{\partial^2 \hat{f}_\alpha}{\partial x^2}-
	E_\alpha\hat{f}_\alpha -	\sigma W_{\alpha\alpha} |\hat{f}_\alpha|^2\hat{f}_\alpha
		 \\ =\hat{V}_{\alpha}\delta(x)\hat{f}_\alpha.
\end{multline}
This means that for even defects we can assume validity of the {\it one-band approximation}  which was the approximation used in Ref.~\cite{ours}. 

The problem of interaction of a soliton of the NLS equation with a localized impurity  has been thoroughly studied in literature (see e.g.  Ref.\cite{Goodman} and references therein). Those results however are not directly applicable to our case, since $\hat{f}_\alpha$ is only an envelope of the WF's while the solution of our problem \eqref{NLS} is given by Eq. (\ref{recover}) [or Eq. (\ref{recover_fin})].

Since the first-band WFs are odd with respect to the spatial variable and the defect potential is localized around the zero of the WFs, for the case of the strong localization $|\hat{V}_{1}|$ is negligibly small compared to $|\hat{V}_{0}|$. Then the defect has much stronger influence on the modes of the zero band  than on those of the first band. Therefore the consideration in this section is restricted to the lowest band, $\alpha=0$. Bright localized modes in this case were considered in Ref.~\cite{ours}. Aiming to extend those studies by including new dynamical properties of the modes in the present subsection we recall the relevant formulas, as well as extend the analysis by including dark modes in the consideration.
  
We will look for  stationary modes in the one-band approximation of the form: $\hat{f}_0(x,t)= e^{-iE t}\phi_0(x)$ where $\phi_0(x)$ is a real function and $E$ is the energy of the mode.  A description and classification of bright defect modes of the NLS equation can be found in Ref.\cite{classif}.  Below we recover some of the known results, which in our case are applicable to the envelope. Besides we will also discuss dark defect modes, as well as the effect of the negative mass and of the fine structure, i.e.  WF's constituting a background of a defect modes, on the mode dynamics.

A relevant parameter of the problem is a detuning of the mode energy outwards the band edge, which is defined as $\varepsilon_0=E-E_0$. 
Then considering $\varepsilon_0\ll E_1-E_0$, what means smallness of the detuning,
and keeping only the terms of the leading order, we can simplify Eqs. (\ref{NLS-del}) reducing them to the stationary NLS equation with a delta impurity
\begin{eqnarray}
	\label{NLS1}
 \frac{1}{2M_0} 
 \frac{d^2 \phi_0}{d x^2}+\varepsilon_0 \phi_0	-
	\sigma W_{00} \phi_0^3
 		=\hat{V}_0\delta(x)\phi_0.  
\end{eqnarray} 
 
Stationary bright modes of Eq. (\ref{NLS1})  can be excited with the energy belonging to the gap, i.e. for a positive detuning: $\varepsilon_0>0$. Introducing $\sigma_V=$sign$(\hat{V}_0)$ and taking into account that $M_0<0$, one finds that the {\it cosh-mode} can be excited only if $\sigma>0$ and in a general case has the form
\begin{eqnarray}
\label{cosh}
\phi_0(x)=\frac{\sqrt{2\varepsilon_0/W_{00}}}{\cosh\left[\sqrt{-2M_0\varepsilon_0}(|x|+x_0)\right]},
\end{eqnarray}
where 
\begin{eqnarray*}
&&x_0=\frac{\sigma_V}{\sqrt{-2M_0\varepsilon_0}}
\mbox{atanh}\sqrt{\frac{\varepsilon_*}{\varepsilon_0}},
 \end{eqnarray*}
with $\varepsilon_0>\varepsilon_*$ and $\varepsilon_*=-\frac{1}{2}M_0\hat{V}_0^2$. 
In the case $\hat V_0>0$ the cosh-mode has one maximum and otherwise it has a two-hump profile (see Fig.2 and the respective discussion in Ref.\cite{ours}).

In the case of attractive interactions, i.e. when $\sigma <0$, and only if $\hat{V}_0>0$ one can find another bright localized solution of Eq. (\ref{NLS1}) -- the {\it sinh-mode} 
\begin{eqnarray}
\label{sinh}
\phi_0(x)=\frac{\sqrt{2\varepsilon_0/W_{00}}}{\sinh\left[\sqrt{-2M_0\varepsilon_0}(|x|+x_0)\right]},
\end{eqnarray}
where
\begin{equation}
x_0=\frac{1}{\sqrt{-2M_0\varepsilon_0}}
\mbox{atanh}\sqrt{\frac{\varepsilon_0}{\varepsilon_*}}.
\end{equation}
and $\varepsilon_0<\varepsilon_*$ [for examples see below Fig.~\ref{fig2} (a)-(c)].

Finally, in case of attractive interactions, $\sigma<0$, considering energy detuning towards the allowed band, i.e. $\varepsilon_0<0$, one can find a {\it dark} localized mode:  
\begin{eqnarray}
\label{tanh}
\phi_0(x)=\sqrt{-\frac{\varepsilon_0}{W_{00}}}\tanh\left[\sqrt{\varepsilon_0 M_0}(|x|+x_0)\right],
\end{eqnarray}
where
\begin{eqnarray}
x_0 =\frac{ \sigma_V}{\sqrt{\varepsilon_0 M_0}}{\rm atanh}\left[  \sqrt{\frac{\varepsilon_*}{2\varepsilon_0}}\left(1-\sqrt{1+\frac{2\varepsilon_0}{\varepsilon_*}}\right)\right]\,.
\end{eqnarray}
  It follows from the obtained formula that
for $\hat V_0<0$ one can construct dark modes  with an envelope having one hole at the defect position [see Fig.~\ref{fig2} (d)-(f)] while in the case  $\hat V_0>0$ the envelope of the dark defect mode has two holes [see Fig.~\ref{fig2} (g)-(i)]. 

\subsection{Numerical results: Comparison with the analytic approximation and stability}

In what follows we will compare the previous analytical results with direct numerical calculations of the defect mode profiles.
To do so we take the same form of the defect potential as in \cite{ours} 
\begin{eqnarray}
\label{defect}
V_d(x)= \frac{\eta}{\sqrt{2\pi}\ell}e^{-\frac{(x-x_d)^2}{2\ell^2}}.
\end{eqnarray}
This defect is characterized by its amplitude $\eta$, width $\ell$ and the location of its center $x_d$. 
Unless otherwise specified we will assume that $x_d$ coincides with the minimum of the potential located at $x = 0$ (this also implies that $A<0$).  
The chosen defect potential can be generated by illuminating the condensate with a localized beam with gaussian profile and orientation perpendicular to the beams generating the OL potential. It is also easy to obtain from Eq. (\ref{defect}) the limit of a point defect (i.e. the delta potential) by taking  $\ell \rightarrow 0$.

As it follows from the list of the hopping integrals $W$ presented in Table~\ref{table1}, for obtaining $W_{00}$ with accuracy of about 3\%, even in the case of amplitude $A=-1$ it is enough to take into account only nonlinear overlapping integrals between neighbouring sites [see the definition (\ref{w})]. 

\begin{widetext}
\begin{figure*}[ht]
\epsfig{file=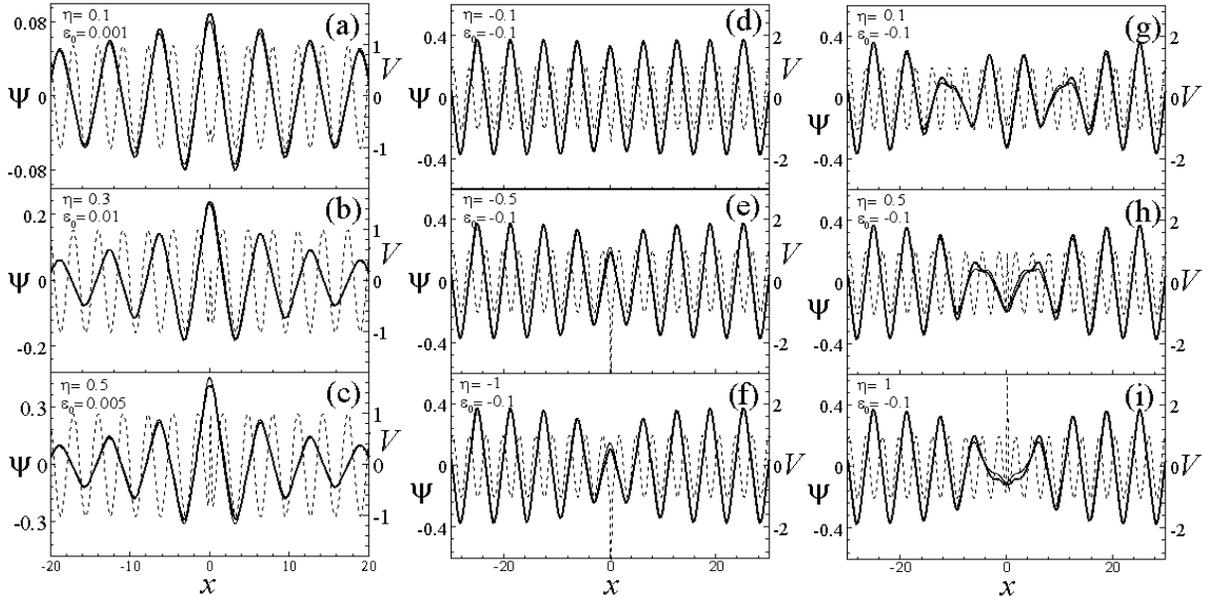,width=16cm}
\caption{Stationary defect modes of the lowest band ($\alpha=0$) for $A=-1$ and $\ell=0.1$. 
Thick  and thin lines correspond to numerical and analytical solutions respectively.
In (a)-(c) we show a sinh-mode for $\sigma=-1$, $V_d>0$,  in (d)-(f) we plot the tanh-mode for $\sigma=-1$, $V_d<0$ and in (g)-(i) two-hole tanh-mode for $\sigma=-1$, $V_d>0$   are presented.
In all the figures $A$, $\ell$ and $\eta$ are given parameters of the lattice and of defect, 
while $\varepsilon_*$, $\hat{V}_0$, and $x_0$ are computed through their definitions given in the text.
The dashed lines show combined potentials whose scale is indicated on the right axes.}
\label{fig2}
\end{figure*}
\end{widetext}

The numerical study of cosh-modes was done in our previous paper 
 \cite{ours}. Here we complete analysis of bright defect modes by presenting in Fig.~\ref{fig2} (a)-(c) sinh-modes calculated analytically from Eqs.~(\ref{recover_fin})  and (\ref{sinh}) and compare them with modes calculated as numerical solutions of Eq.~(\ref{NLS}) for different defect parameters.
Considering the lowest band, $\alpha=0$, we find that sinh-modes exist only when $V_d>0$ and  $\varepsilon_0< \varepsilon_*$. 
In this case, the defect should be taken sufficiently large to ensure that $\varepsilon_*$ differs appreciably from zero.

In Fig.~\ref{fig2} (d)-(i) we present dark-modes in the form of tanh-function calculated analytically from Eqs.~(\ref{recover_fin})  and (\ref{tanh}). Dark modes have energy shift $\varepsilon_0$ towards the allowed zone and exist at the boundary of the BZ only for attractive nonlinearity, $\sigma<0$. We also observe that dark modes exist for any ratio $\varepsilon_0/\varepsilon_*$.
For negative $V_d$ one can construct one-hole tanh-mode [see Fig.~\ref{fig2} (d)-(f)] and for positive $V_d$ tanh-modes with two holes [Fig.~\ref{fig2} (g)-(i)]. The holes however refer to the envelope of the WFs and therefore are not always visible in the profile, as it happens for instance, in Fig.~\ref{fig2} (h), (i).

We have checked using analytical formulas as well as direct numerical simulations that by decreasing the defect width by factor of ten, the shapes of defect modes do not change 
appreciably and their amplitudes differ only by $1\%$, what confirms the good accuracy obtained from the delta-approximation for all $\ell <\pi$. 
Moreover, Fig.~\ref{fig2} shows, that in the outlined parameter region, our analytical approach gives the mode profiles with sufficiently high accuracy, even in the one-band approximation. 
Obviously, increasing the value of the defect strength, $\eta$, the analytical approximation becomes worse.  When $\eta$ becomes of order one, the difference between the analytical predictions and the numerical results becomes appreciable. This occurs because such a defect originates modes localized on very few lattice periods and the effective mass approach~\cite{KS} fails. Then one expects that the tight-binding approximation~\cite{AKKS,KAKS} would be more appropriate \cite{PP}.

\begin{widetext}
\begin{figure*}[ht]
\epsfig{file=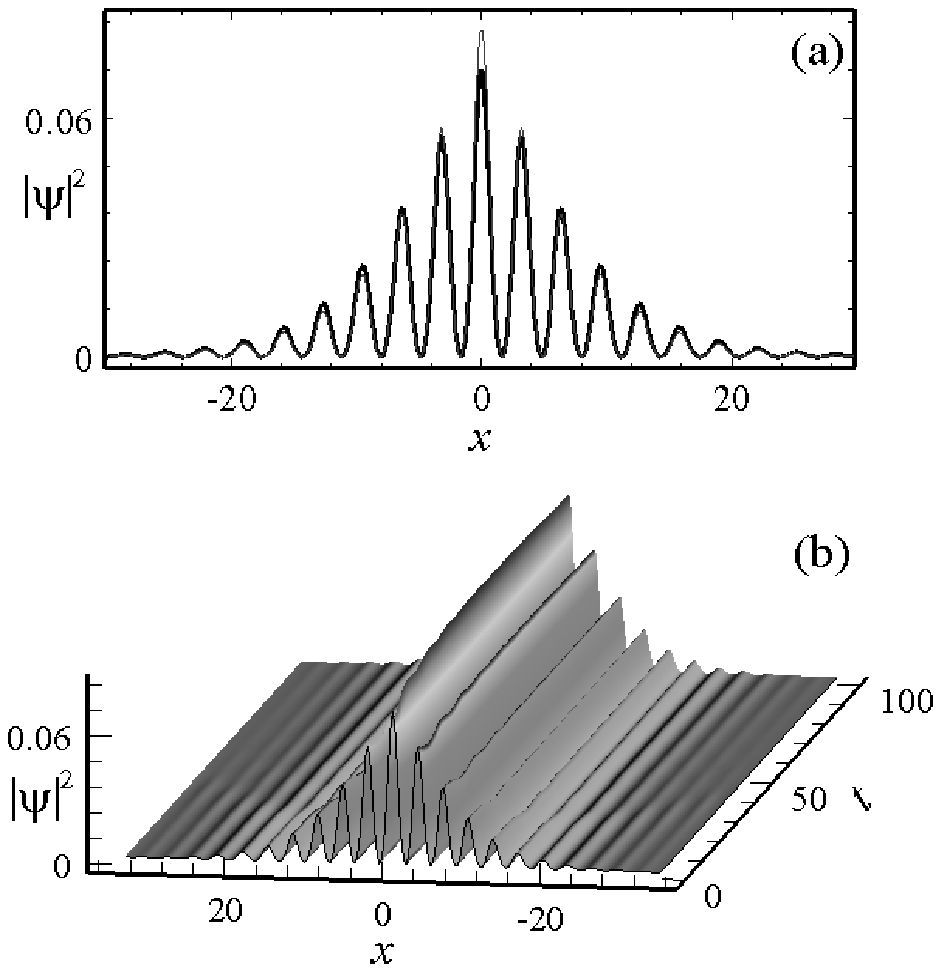,width=3.5cm}
\epsfig{file=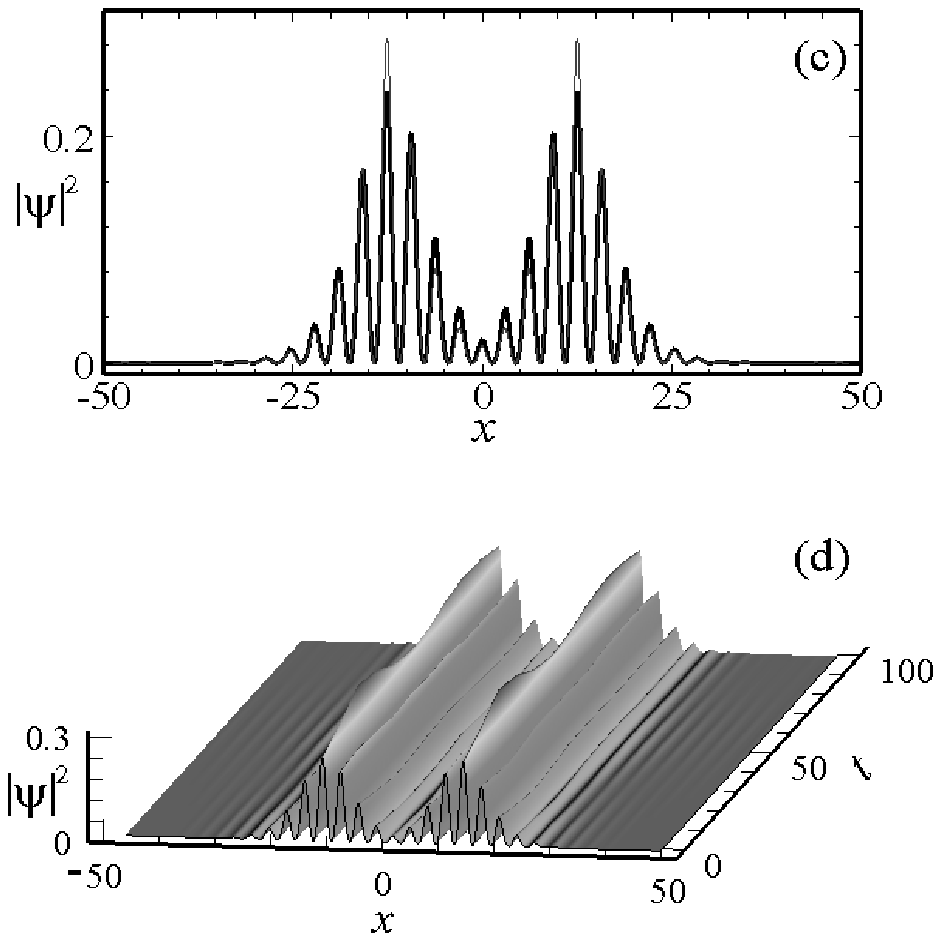,width=3.5cm}
\epsfig{file=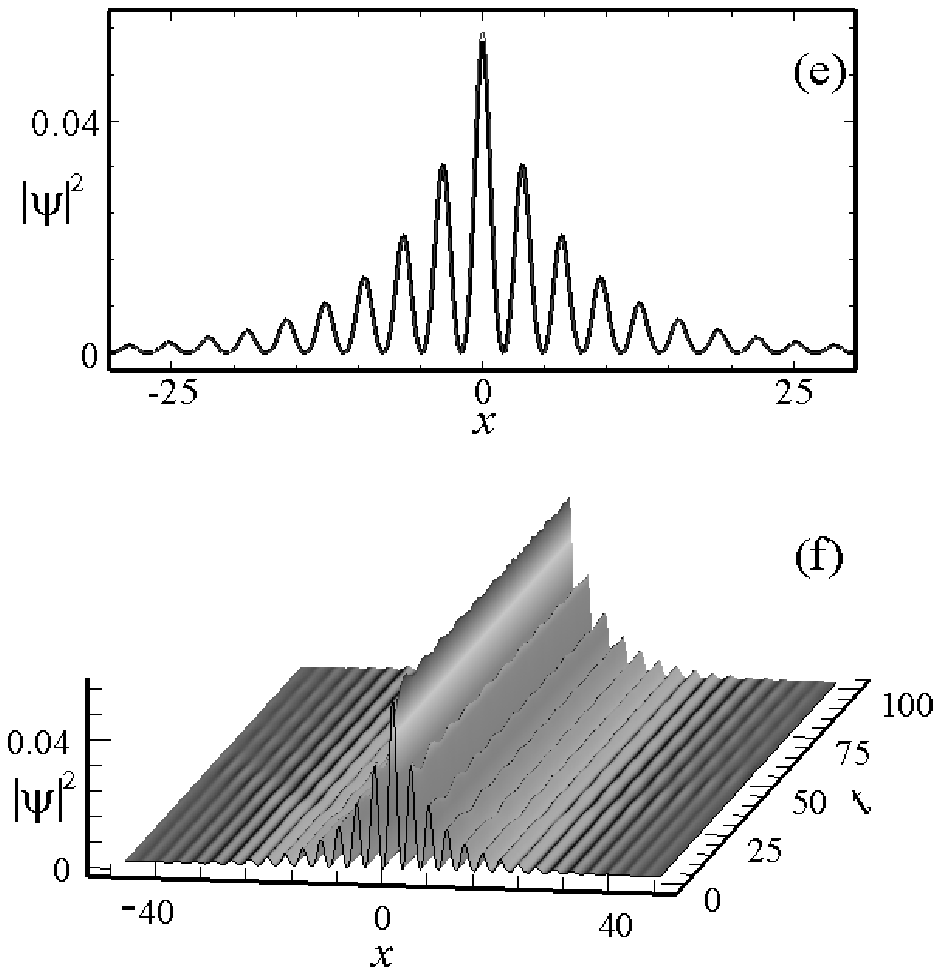,width=3.5cm}
\epsfig{file=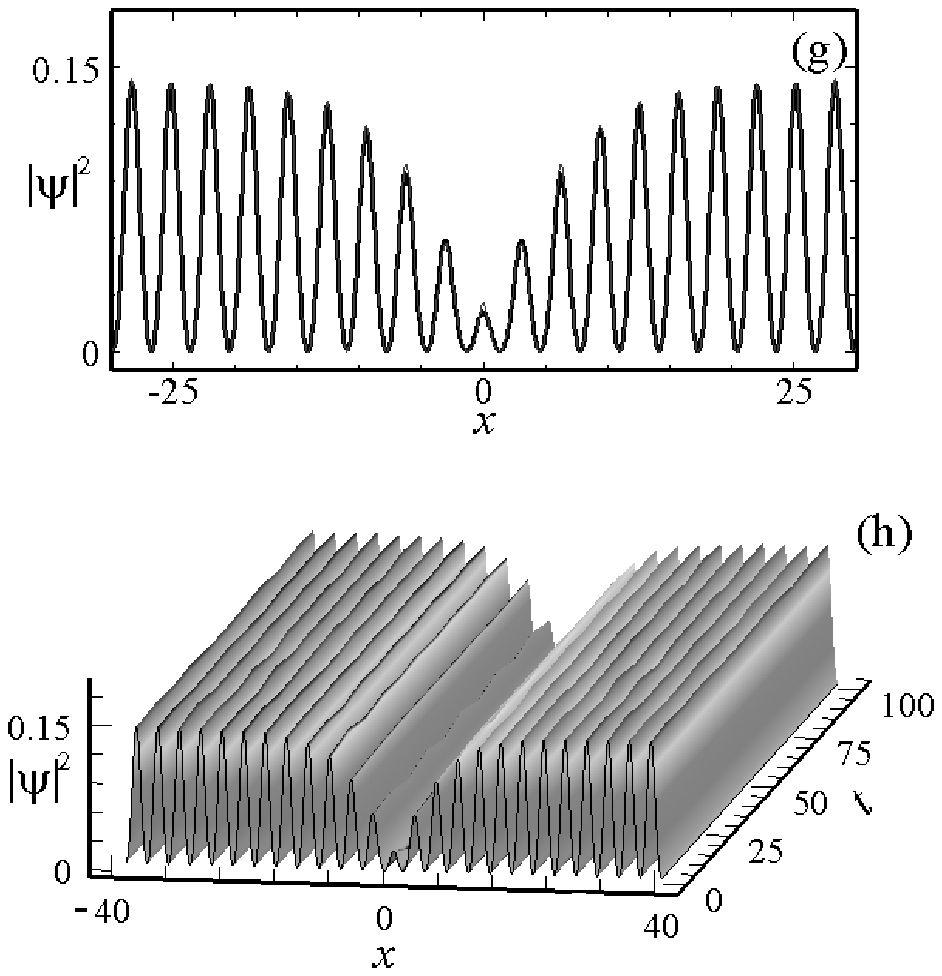,width=3.5cm}
\epsfig{file=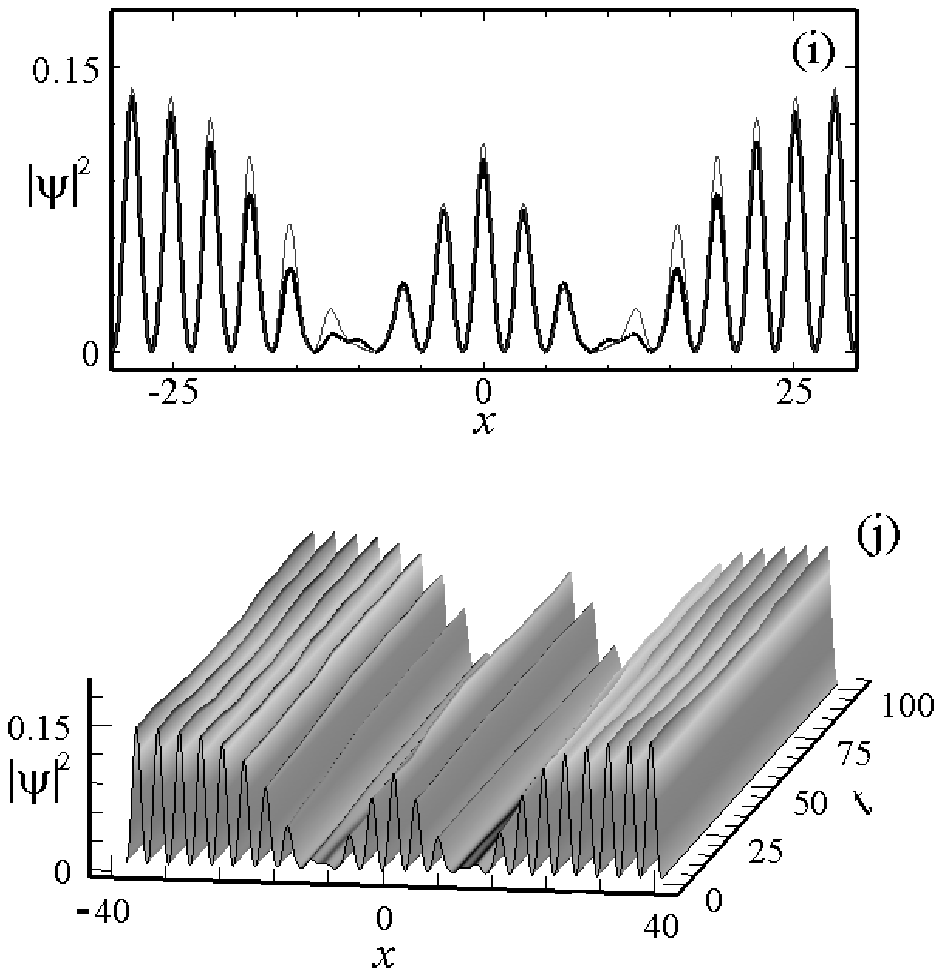,width=3.5cm}
\caption{In (a), (c), (e), (g) and (i) the initial (thick, black) and final (thin, red) profiles of the defect modes and in (b), (d), (f), (h) and (j)  their corresponding dynamics are shown. The initial profiles for cosh-modes is calculated by using the following parameters: $A=-1$, $\sigma=1$, $\ell=0.1$ and $\eta=0.1$, $\varepsilon_0=0.03$ for (a) and $\eta=-0.5$, $\varepsilon_0=0.09$ for (c).
The initial profiles in  (e), (g) and (i) are the same as the analytical solutions depicted in Fig. \ref{fig2} (b) (f), (g), correspondingly.}
\label{fig3}
\end{figure*}
\end{widetext}

The stability of the defect modes was tested by computing the evolution of the cosh-, sinh- and tanh-modes, starting from their analytical expressions and shown in Fig.~\ref{fig3} (b), (d), (f), (h) and (j).  Since the respective data  are not exact solutions of Eq.~(\ref{NLS}), such initial conditions automatically introduce perturbations of order of a few percents with respect to the true stationary mode.  
In all the cases depicted in  Fig.~\ref{fig3} one observes stable evolution of the modes.     

\subsection{Coupled modes for an odd defect}

Coupling of the bands can occur either due to the nonlinearity or due to antisymmetric part of the defect, $V_-$. In the present subsection we concentrate on the last possibility, while still neglecting the cross-phase modulation. More specifically, we consider stationary solutions
$\hat{f}_\alpha(x,t)=e^{-iEt}\phi_\alpha(x)$, for which the system (\ref{NLS-del}) is reduced to  
\begin{subequations}
	\label{NLS-del-1}
\begin{eqnarray}
	\frac{1}{2M_0} \frac{\partial^2 \phi_0}{\partial x^2}+\varepsilon_0\phi_0 -	\sigma W_{00} \phi_0^3=  
\hat{V}\delta(x)\phi_1\,,
\\
\frac{1}{2M_1} \frac{\partial^2 \phi_1}{\partial x^2}+\varepsilon_1\phi_1- 	\sigma W_{11} \phi_1^3 = \hat{V}\delta(x)\phi_0\,. 
\end{eqnarray}
\end{subequations}

For obtaining a stationary localized nonlinear defect mode of (\ref{NLS-del-1}), one has to take into account that in the case at hand $M_0M_1<0$. Then solving the system (\ref{NLS-del-1}) one can compute
\begin{subequations}
\label{odd_mod}
\begin{eqnarray}
\label{odd_mod0}
&&\phi_\alpha(x)=\frac{\sqrt{2|\varepsilon_\alpha|/W_{\alpha\alpha}}}{\cosh\left[b_\alpha(|x|-x_\alpha)\right]}
\\
&&\phi_\beta(x)=\frac{\sqrt{2|\varepsilon_\beta|/W_{\beta\beta}}}{\sinh\left[b_\beta(|x|-x_\beta)\right]}
\label{odd_mod1}
\end{eqnarray}
\end{subequations}
where $b_{\gamma}=\sqrt{2|M_{\gamma}\varepsilon_{\gamma}|}$, the indexes $(\alpha,\beta)$ are $(0,1)$ for $\sigma>0$ or $(1,0)$ for $\sigma<0$, correspondingly, and the constrain $M_\alpha \varepsilon_\beta>0$ is verified. The parameters
$x_{\alpha,\beta}$ are found from the following system
\begin{subequations}
\label{x_odd_mod}
\begin{eqnarray}  \tanh(b_\alpha x_\alpha)\frac{\sinh(b_\beta x_\beta)}{\cosh(b_\alpha x_\alpha)}&=&   -\frac{\hat{V}}{\varepsilon_\alpha}\sqrt{\frac{\varepsilon_\beta M_\alpha W_{\alpha\alpha}}{2W_{\beta \beta}}}, \\
  \frac1{\tanh(b_\beta x_\beta)}\frac{\cosh(b_\alpha x_\alpha)}{\sinh(b_\beta x_\beta)}&=&   \frac{\hat{V}}{\varepsilon_\beta}\sqrt{\frac{\varepsilon_\alpha M_\beta W_{\beta\beta}}{2W_{\alpha \alpha}}}.
\end{eqnarray}
\end{subequations}

Strictly speaking, mathematical consistency of the obtained system of equations requires smallness of both detuning parameters $\varepsilon_\alpha$ and $\varepsilon_\beta$. 
This follows from the imposed condition of the smoothness of the envelopes $\phi_{\alpha,\beta}$ compared to the lattice constant (what allowed us to introduce the effective masses). 
That is why, the applicability of the two-level approximation is determined by the width of the gap, $\Delta=|\varepsilon_\alpha|+|\varepsilon_\beta|$, which should be small, in order that both detunings $|\varepsilon_{\alpha,\beta}|$ be small. This can be provided by small amplitude OLs. On the other hand one should have WFs localized on a few periods of the OL what can be satisfied by increasing the amplitude of the OL. These arguments make applications of the system (\ref{odd_mod}) to be rather restricted and requires a numerical verification of the obtained solutions. 

\begin{figure}[ht]
\epsfig{file=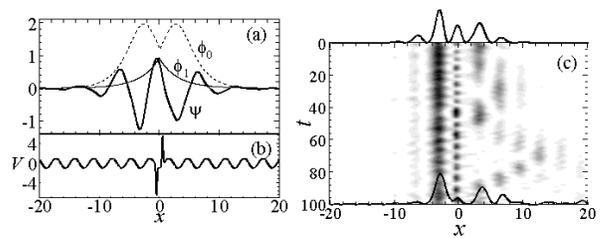,width=8cm}
\caption{(a) Stationary coupled mode $\psi(x,0)$ for $\alpha=0$, $\beta=1$ and the corresponding envelopes $\phi_0$ and $\phi_1$ are shown. 
(b) Combined potential $V(x)=V_l(x)+V_-(x)$.
The parameters are: $A=-1$, $\eta=1.5$, $\ell=0.1$, $x_d=0.5$, $E=0.9688$ ($\varepsilon_0=-\varepsilon_1=0.4982$). Some parameters such as $W_{\alpha\alpha}$, $M_\alpha$ are taken from Table I.
(c) Time evolution of the coupled mode shown in (a). The solid black lines indicate the initial (top) and final (bottom) profiles.}
\label{oddmodes}
\end{figure}

To this end we choose the odd defect in the following form
\begin{eqnarray}
V_d(x)=V_-(x)= \frac{\eta}{\sqrt{2\pi}\ell}\left(e^{-\frac{(x-x_d)^2}{2\ell^2}}-e^{-\frac{(x+x_d)^2}{2\ell^2}}\right)
\end{eqnarray}
which according to Eq.(\ref{V_approx_b}) gives only antisymmetric component $\hat V$ of the defect 
while symmetric components $\hat V_0$ and $\hat V_1$ are zero. Next we recall the results presented in Table~\ref{table1} where the effective masses for the potentials $A=-1$ were found to be of order of $0.1$ while the width of the lowest gap is approximately 1. Thus, one can expect that in this case the above requirement are approximately satisfied.  In Fig.~\ref{oddmodes} we show an example of a defect mode, which first was computed using the analytical approximation from the expressions (\ref{odd_mod0}), (\ref{odd_mod1}) and (\ref{recover_fin}) (i.e. the expansion over the WFs [see Fig.~\ref{oddmodes} (a)] in the potential shown in the panel (b). Next, the constructed mode was considered as an initial condition for Eq. (\ref{NLS}) and was allowed to evolve freely in time. The evolution is shown in Fig.~\ref{oddmodes} (c).

In Fig.~\ref{oddmodes} (c) one can observe relatively stable dynamics, although accompanied by emission of some radiation, what means that the analytic approximation is relatively good. Meantime, any significant deviation of the potential amplitude from the unity, has led  to unstable dynamics of the initial conditions obtained from (\ref{odd_mod0}), (\ref{odd_mod1}) and (\ref{recover_fin}), what is natural in view of the limitations of the theory described above. 

\section{Driving defect modes through the optical lattice}
\label{IV}

\subsection{Response of defect modes to a simple shift of the defect position}

\begin{figure}[h]
\epsfig{file=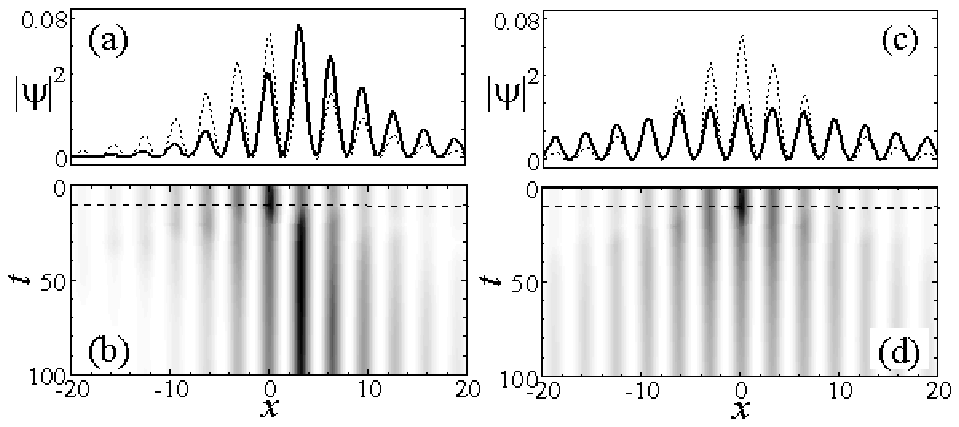,width=6.5cm}
\epsfig{file=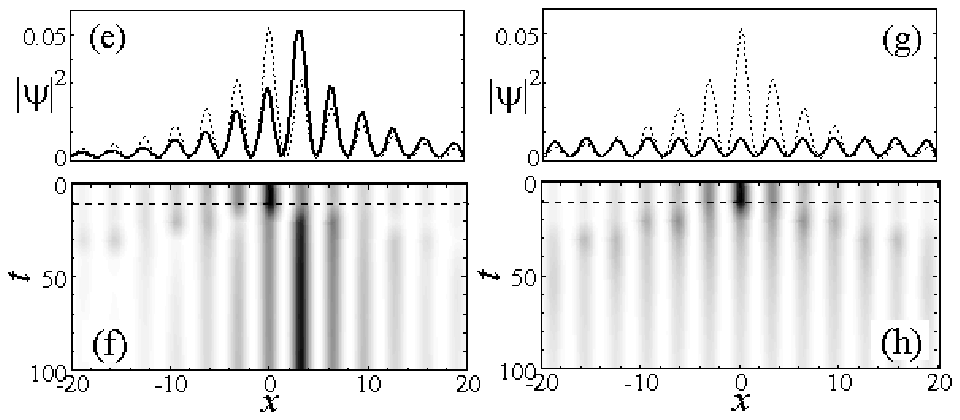,width=6.5cm}
\epsfig{file=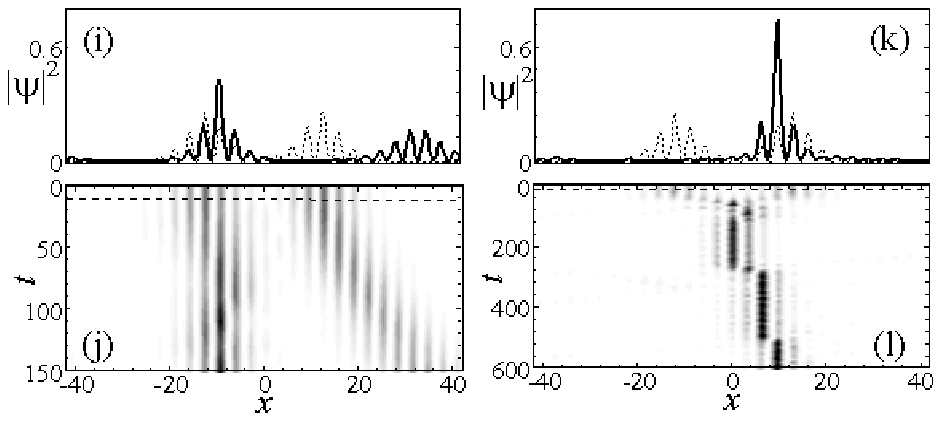,width=6.5cm}
\epsfig{file=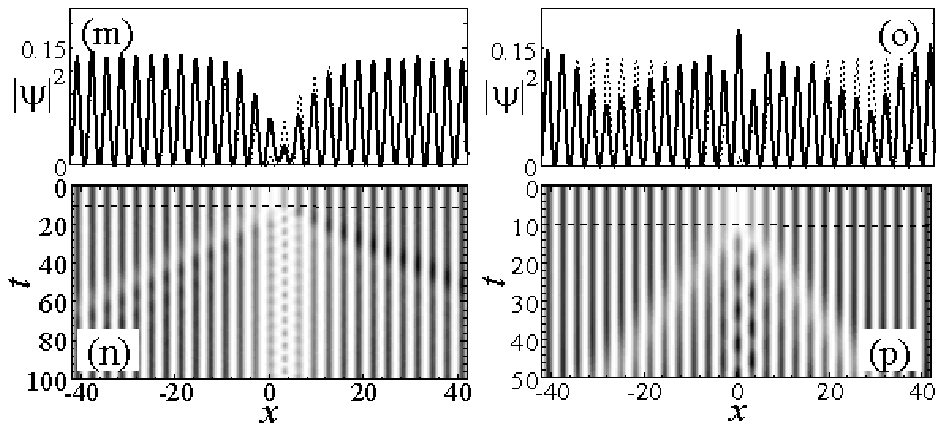,width=6.5cm}
\epsfig{file=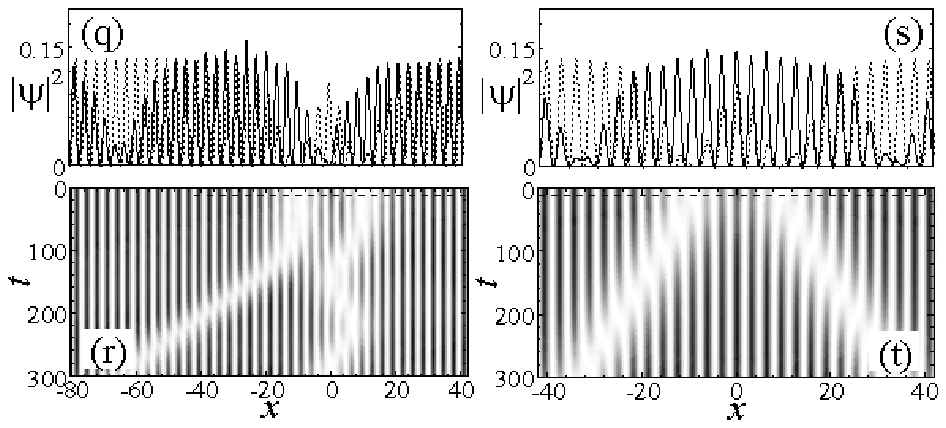,width=6.5cm}
\caption{ Dynamics of the defect modes under defect shifts in the positive direction at time $t=10$ by $\pi$ (left column) and by $\pi/2$ (right column).  The blocks of four panels from the top correspond to one-hump cosh-mode (a)-(d); sinh-mode (e)-(h); two-hump cosh-mode (i)-(l); one-hole tanh-mode (m)-(p); and two-hole tanh-mode (q)-(t).   Initial profiles (dashed lines) 
and defect parameters in (a,c), (e,g), (i,k), (m,o) and (q,s) are the same as those in Fig. \ref{fig3} (a), (c), (e), (g) and (i), correspondingly. The final profiles (solid lines) are taken from the final time of the corresponding density plots. 
In the panels with density plots the horizontal dashed line shows the time of the shift of the defect.}
\label{fig_shift}
\end{figure}

>From the physical point of view and also thinking on possible applications of defect modes, 
it is relevant to understand if they can be driven through the lattice and what are the limitations of those motions. Following the analysis of Ref.~\cite{ours}, here we start by studying the behavior of  defect modes under a shift of the defect position. We first consider evolution of the simplest cosh- and sinh- modes affected by the shift of the defect center by a lattice period (i.e. by $\pi$ in our case). As one can see from Fig.~\ref{fig_shift} (a), (b) (for cosh-mode) and Fig.~\ref{fig_shift} (e), (f) (for sinh-mode), in this situation, the defect modes follow the position of the center of the defect potential, i.e. after some relaxation time  the defect mode follows the defect position. 
However, when the defect is shifted by a half-period, both modes become unstable and spread out as it is shown in Fig.~\ref{fig_shift} (c), (d) (for cosh-mode) and Fig.~\ref{fig_shift} (g), (h) (for sinh-mode). The observed behavior can be understood by realizing that a positive defect corresponds to local repulsion of the atoms having real, and thus positive, mass. Then, being placed on top of the maximum of the lattice potential the defect results in strong local repulsion of the atoms, what leads to decay of the defect mode. This does not happen when the shift of the defect mode is $\pi$ since in that case the local repulsion induced by the defect is compensated by the local attraction of the lattice potential, and for the whole excitation the effective mass approximation is applied. 

In the case of a two-hump cosh-mode the picture changes drastically.
By shifting the defect to the next adjacent minimum (in our case to the right) we observe that the mode breaks into two parts: a one-hump mode, which after a relaxation time stabilizes around the new potential minimum where the defect is centered, and a mode moving outwards the defect with a constant  velocity [see Fig. \ref{fig_shift} (i)-(j)]. 
If, however, the defect is shifted to the nearest potential maximum (i.e. to $x_d=\pi/2$) the two-hump cosh-mode evolves into a one-hump mode stabilized at the position of the defect [Fig. \ref{fig_shift}~(k), (l)]. 
While this behavior may seem contradictory to what we observed in the case of the one-hump cosh-mode,  it can be understood by taking into account that now the defect is negative:  $\hat{V}_0<0$, and thus acts as repulsive impurity for the wavepackets having negative effective mass and as attractive impurity for real atoms. 

Let us now think of the two-hump solutions as two wavepackets whose attraction is balanced by the repulsive defect placed between them. Then, a shift of the defect to the right by one lattice period destabilizes this balance and pushes the right-placed wavepacket in the positive direction, what leads to its motion [see Fig.~\ref{fig_shift}~(i),(j)]. Also, the attraction between the two wavepackets results in a small shift of the left-placed wavepacket toward the defect. When  the defect is shifted by $\pi/2$ and is placed on top of the nearest local maximum of the OL [see Fig. \ref{fig_shift} (k),(l)], its effect is dramatically reduced by the lattice maximum, which acts in the opposite sense. Therefore, the repulsion between the left and right waves is reduced what leads to their motion toward each other and collapsing in one localized wavepacket. The proposed interpretation of the phenomenon was supported by further studies of the dynamics of two-hump cosh-modes. First, after a sudden switch off of the defect without any dislocation, we observed a merging of the two wavepackets very similar to one shown in Fig.~\ref{fig_shift}~(l). Next, we have observed continuing motion of the wavepacket (after separation with the defect, it can be more appropriately called envelope soliton) as it would be reflected by the defect, although weakened by the lattice potential. The first stages of such dynamics are shown in Fig.~\ref{fig_shift}~(l).   Finally, by increasing the strength of the defect we succeeded to create a more  localized wavepacket moving faster.


The response of the tanh-modes, given by the formula (\ref{tanh}), under a shift of the defect is shown in Fig.\ref{fig_shift}~(m)-(t). The one-hole tanh-mode follows the defect shift to the nearest potential minimum accompanied by excitation of oscillating dynamics of the mode amplitude. When the defect is shifted to the nearest maxima the one-hole tanh-mode  breaks into two modes which move outwards  in opposite directions [see Fig.~\ref{fig_shift}~(o),~(p)]. In order to understand the dynamics of a dark mode, in the same spirit as this was done above for the bright modes, we interpret the mode as a dark soliton interacting with a delta-impurity and recall the results on such interaction~\cite{KPTV}. In the case at hand $\hat{V}_0<0$ and $M_0<0$, what in terms of the equation (\ref{NLS-del-sinlge}) means that the impurity attracts the soliton having initially zero velocity: as it is shown in Ref.~\cite{KPTV} [see the formula (16) there] the initial acceleration of the soliton due to attraction is proportional to $M_0\hat{V}_0$. This explains the dynamics shown in Fig.~\ref{fig_shift}~(m),~(n). When the defect has a $\pi/2$ shift, as it was explained above, its strength in the final position is significantly weakened (it is now placed on the top of a lattice maximum).  
Since the shift of the defect acts as a perturbation of the initials conditions which leads to the decay of a static soliton into two solitons with finite velocities [moving outwards the defect according to the momentum conservation law, as it is shown in Fig.~\ref{fig_shift}~(o),~(p)].

The decay of defect modes into two solitons is more evident in the case of a two-hole initial condition  and with a repulsive impurity which is repulsive for the static soliton (see Eq. (16) in Ref.~\cite{KPTV}) shown in Fig.\ref{fig_shift} (s), (t) for the $\pi/2$ shift of the defect position. At the same time, the  two-hole tanh-mode displays more sophisticated evolution, shown in the panel (q), (r) of Fig.~\ref{fig_shift}. After the shift of the defect to the nearest lattice  minimum, the two holes start to behave as two dark solitons, one of them (the one closer to the new position of the impurity) being trapped  while the other one moves far from the impurity.  

\subsection{Driving the defect modes}

In Ref. \cite{ours} it was shown that the robustness of the cosh-mode described in the preceding subsection allows one to drive the defect mode along the lattice. Since the  defect mode is destroyed when it  stays close to the local maxima of the potential, in order to provide stable motion of the mode, the defect motion cannot be uniform, but described by a step-like function which consists of two characteristic time intervals: a fast enough one,  $\tau$, in which the defect is shifted by one lattice period, and another larger one, $T$, allowing for the defect-mode to recover its shape on the next lattice site. 

The physical reasons for the existence of these two characteristic times, $\tau$ and $T$ are rather simple. 
Indeed, the respective microscopic dynamics can be interpreted as a tunneling of part of atoms, attracted by the shifted defect, from the potential well $x=0$ to the neighbor potential well $x=\pi$. 
Let the characteristic tunneling time be $\tau_0$. Taking into account that all intermediate states (i.e. those with center different from $\pi n$, where $n$ is an integer) are unstable, one finds that the requirement $\tau\ll \tau_0 \ll T$ must be satisfied. This last constrain evidently limits the speed with which the mode can be driven along the lattice.

In order to find a numerical estimate for the tunneling time $\tau_0$ we take into account that we are working in the framework of the effective mass approximation which implies relatively small differences in populations of neighbor potential wells. 
Also we are dealing with the modes bordering the boundary of the BZ, what means that the phase difference between atoms in adjacent cells is $\pi$. Thus we can use the results for the half-period of oscillations of the linear tunneling of condensate in a double-well potential \cite{RSFS}:
\begin{equation}
	\label{tun-time}
	\tau_0=\frac{\pi}{2\sqrt{K^2-W_{0000}^{0000}K}},
\end{equation}
 with
\begin{equation}
	K=-\int\left[\frac{dw_{00}}{dx}\frac{dw_{01}}{dx} +Aw_{00}w_{01}\cos(2x)\right]dx,
\end{equation}
for a particular potential given by (\ref{PerPot})  (here we have taken into account that the WFs make up an orthonormal basis). In particular, for $A=-1$, the mobility conditions are that the defect must be shifted with 
 a velocity  much bigger than 
 \begin{equation}
 \label{ma}
 v_* = \frac{\pi}{3.67} \approx 0.86,
 \end{equation}
  and then leave the system to relax for a time 
\begin{equation}\label{mb}
  T\gg \tau_0 \approx 3.67.
 \end{equation}
 
\begin{figure}[th]
\epsfig{file=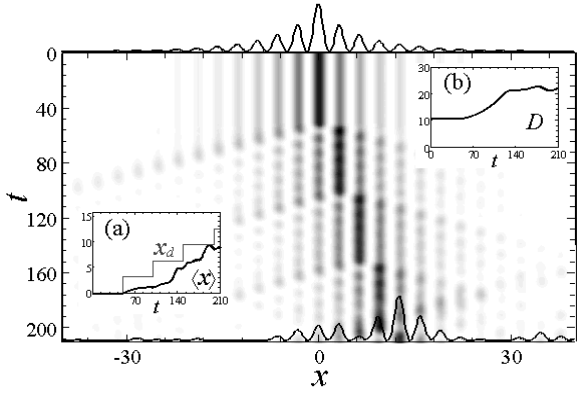,width=8cm}
\caption{ Density plot of the sinh-mode having the initial parameters as in Fig. \ref{fig3} (c) driven by the moving defect following trajectory $x_d(t)=\pi\sum_{j=0}\theta (t-T\,j)$ with $T=50$. 
The solid black lines indicate the initial (top) and final (bottom) profiles. 
In the inset (a)  the dynamics of the center of mass $\langle x\rangle$ of the wavepacket   
(thick black line) and of the defect position $x_d(t)$ (thin blue line) are shown. 
In inset (b) the dynamics of the wavepacket dispersion $D(t)$ is presented.
}
\label{figmove}
\end{figure}

In  Ref. \cite{ours} (see Fig.4) we have presented an example of how the defect cosh-mode can be driven along the lattice by moving the position of the defect. 
According to the properties shown Fig. \ref{fig_shift} (e), (f), the sinh-mode also can be moved through the lattice. An example is shown in Fig.\ref{figmove}, where we present the dynamics of the mode as well as the trajectory of the  center of the wavepacket $\langle x\rangle$ 
 \begin{eqnarray}
\langle x\rangle=\frac 1N\int x|\psi|^2dx 
\end{eqnarray}
and of its dispersion
\begin{eqnarray}
\quad D=\sqrt{\frac 1N\int (x-\langle x\rangle)^2|\psi|^2dx}\,.
\end{eqnarray}
Here as usually
$
	N =\int |\psi|^2dx
$
is a number of particles.

Due to the higher scattering on the potential barriers after each defect, shifting the dynamics of this mode is less robust than 
for the cosh-mode what leads to its faster dissipation as shown in  Fig.\ref{figmove}. 
Also, due to significant reflection on each period of reconstruction one observes larger deviation of the center of mass of the mode $\left\langle x \right\rangle(t)$ from the defect trajectory $x_d(t)$ as well as significant growth of the width of the wavepacket $D$ [see Fig.\ref{figmove} (a), (b)].
Therefore in the following analysis of the motion of defect modes through the lattice we will use cosh-mode.

\begin{figure}[ht]
\epsfig{file=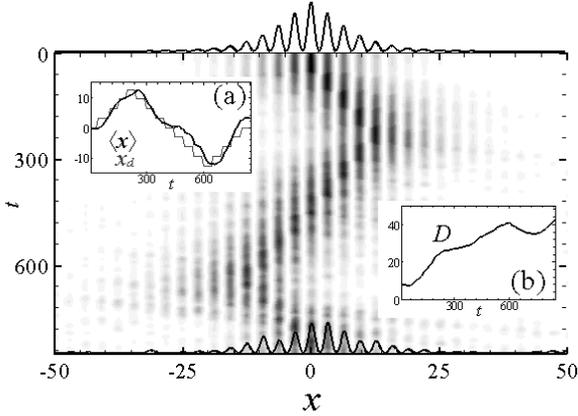,width=8cm}
\caption{ Density plot of driving of a cosh-mode with initial parameters as in Fig. \ref{fig3} (a) 
by moving the defect following a zigzag trajectory according to the law $x_d(t)=\pi\sum_{j=0}\theta (t-t_0-T\,j)$ along each direct interval with $T=50$. 
At times $t=250$ and $650$ we invert the direction of the motion of the defect which is followed by the inversion of the direction of the motion of the mode.
The solid black lines indicate the initial (top) and final (bottom) profiles. 
In (a) dynamics of the center of mass $\langle x\rangle(t)$ 
of the wavepacket (thick black) is shown. The defect position $x_d(t)$ is indicated by the thin blue line in the same plot. 
Inset (b) presents the dynamics of the wavepacket width $D(t)$.
}
\label{figmove_ZZ}
\end{figure}

If the motion of the defect satisfies the conditions  \eqref{ma} and \eqref{mb}, more sophisticated dynamics can be generated by moving defects. For instance, in Fig.~\ref{figmove_ZZ} we show a zigzag trajectory of the cosh-mode induced by the corresponding zigzag motion of the defect.
The defect mode experiences some changes when driven through the lattice. For instance, when driven over ten lattice periods the amplitudes of the initial and final maxima differ by about 10\%.

\begin{figure}[ht]
\epsfig{file=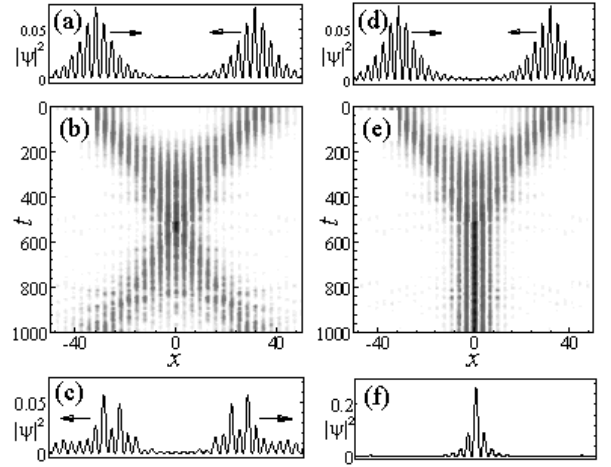,width=8cm}
\caption{
Two regimes of collision of two identical cosh-modes, each having the initial parameters as in Fig. \ref{fig3} (a). The initial distance between the modes is 20 periods. 
In both columns initially the modes are driven by the defects toward each other according to the law used in Fig. \ref{figmove} and \ref{figmove_ZZ} with $T=50$ what results in a collision at $t=500$ (the time where the defect positions coincide). In the left column we show the dynamics where after collision defects continue to drive modes along the initial directions. In the right column the results correspond to the situation when the defects are stopped after they are met. Panels (b) and (e) show the respective temporal dynamics, while panels (c) and (f) show the final outputs at $t=10^3$.}
\label{figmove_coll}
\end{figure}

We have also carried out  numerical experiments on collisions of defect modes driven by counter-propagating defects. In Fig. \ref{figmove_coll} we show an example with two cosh-modes each of them shifted by ten periods from the center in opposite directions.
By moving the defects towards each other we induce a head-on collision at $t = 500$. When the defects continue to be driven after the collision [Fig. \ref{figmove_coll}(a)-(c)]
they pass through each other and emerge with substantial shape modifications but still being identifiable [Fig. \ref{figmove_coll}(c)]. When the defects are stopped at the collision time 
$t \simeq 500$ a single defect mode with larger amplitude and stronger localization appears [Fig. \ref{figmove_coll}( e), (f)].

\section{Generation of defect modes}
\label{V}

\begin{figure}[h]
\epsfig{file=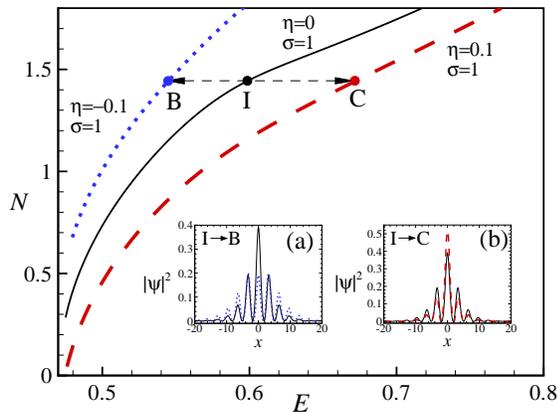,width=8cm}
\caption{
[Color online] Adiabatic excitation of cosh- and two-hump cosh-modes is shown for $A=-1$, $\sigma=1$ and $\ell=0.1$.
Different branches $N(E)$ in the first gap correspond to the lowest gap soliton for $\eta=0$ (black, solid line), cosh-mode for $\eta=0.1$ (red, dashed line) and two-hump cosh-mode for $\eta=-0.1$ (blue, dotted line).
In (a) and (b) the initial profile of the gap soliton (black, solid line) corresponds to the point I and final profiles of the two-hump cosh- and cosh-modes (blue dotted and red dashed, thick lines) in the points B and C, correspondingly, are shown which coincide with the profiles of the defect modes obtained by adiabatic growth of the defect amplitude.}
\label{figexcit_ch}
\end{figure}

\begin{figure}[h]
\epsfig{file=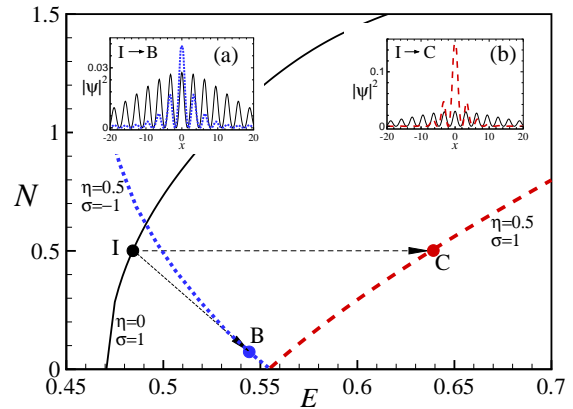,width=8cm}
\caption{[Color online]
Adiabatic excitation of a sinh- and cosh-mode mode is shown for $A=-1$, $\eta=0.5$ and $\ell=0.1$.
Different branches $N(E)$ in the first gap correspond to the lowest gap soliton for $\eta=0$ and $\sigma=1$ (black, thin line), cosh-mode for $\eta=0.5$ and $\sigma=1$ (red, dashed line) and sinh-mode for $\eta=0.5$ and $\sigma=-1$ (blue, dotted line).
In (a) and (b) the initial profile of the gap soliton (black, solid line) corresponds to the point I and final profiles of the sinh- and cosh-modes (blue, dotted and red, dashed thick lines) in the points B and C, correspondingly, are shown which coincide with profiles of the defect modes obtained by adiabatic growth of the defect amplitude. }
\label{figexcit_sh}
\end{figure}

A relevant question for experimental studies of defect modes is their generation in real situations.
One of the possibilities consists in generating, first, a gap soliton with energy near a gap edge  
in an homogeneous OL and subsequently adiabatically switching on the defect by changing the defect amplitude from zero to some value. This  will lead to the transformation of a gap soliton into a defect mode.  A numerical implementation of this scheme is illustrated in Figs. \ref{figexcit_ch} and \ref{figexcit_sh}, where we have presented the dependence of the number of particles $N$ {\it vs} the energy of the mode $E$ for the original gap soliton (thin solid lines) and the emerging defect modes (thick dashed and dotted lines), found by searching stationary modes of Eq. (\ref{NLS}).

More specifically, in order to generate cosh- and two-hump cosh-modes we start with the envelope soliton (as an example  consider, say, the point I in Fig.\ref{figexcit_ch}) with energy $E=E_0+\varepsilon_0$. Next, by increasing $\eta$ [the defect shape is chosen to be (\ref{defect})] we generate one-hump cosh-mode, shown as 
I$\to$C transition [in Fig. \ref{figexcit_ch} (b) and \ref{figexcit_sh} (b)], and by decreasing $\eta$ (i.e. increasing $|\eta|$) we generate a two- hump cosh-mode, shown as (I$\to$B) transition. In both cases we used the stationary 1D GP equation with a periodic potential to compute numerically the stationary solutions and then 
solving the time-dependent equation (\ref{NLS}) with adiabatic growth of the defect amplitude according to the law $\eta =\pm 0.1 \sin^2(10^{-3}\pi t)$. It worth pointing out that in the described transitions the numbers of particles was preserved.

We have followed the same procedure for the generation of a sinh-mode as schematically shown in Fig. \ref{figexcit_sh}.
However, the difference from the previous case is  that to generate a sinh-mode from the gap soliton 
not only the defect amplitude must be increased adiabatically but also the sign of the interparticle interactions, $\sigma$, must be changed from repulsive to attractive. 
Due to the fact that gap soliton does not exists for attractive interaction it starts to spread and the only small number of particles transforms into the sinh-mode  [see the process I$\to$B in Fig. \ref{figexcit_sh}]. 
The resulting number of particles and corresponding profile of the sinh-mode slightly depends on the velocity of growth of the defect amplitude 
[in the present example we used $\eta =0.5 \sin^2(5\times 10^{-4}\pi t)$].

\section{Conclusions}
\label{VI}

In this paper we have investigated localized defect modes of the one-dimensional Gross-Pitaevskii equation in the presence of an optical lattice and a localized defect. 

 Using the expansion of the condensate wave function over Wannier functions it was shown that the problem can be reduced to the study of the dynamics of the envelope which is governed by the nonlinear Schr\"{o}dinger equation with a delta-impurity (in the one-band approximation), or in a general case by a system of coupled nonlinear Schr\"{o}dinger equations with delta-impurities (in the two-band approximation). 
The symmetric analytical solutions the obtained equation in the stationary case describe envelopes of the stationary defect modes of $\cosh-$, $\sinh-$, and $\tanh-$types. Our approximate analytical formulas for impurity modes are in good agreement with results obtained by direct numerical calculations.

We have also verified the stability of the impurity modes by direct numerical simulations of the underlying one-dimensional Gross-Pitaevskii equation with a periodic potential and a localized defect showing that the defect modes are indeed stable under small perturbations (i.e. linearly stable).

Using the same ideas developed in Ref. \cite{ours}  we have shown that impurity cosh-modes can be made to follow complex trajectories of the defect position while sinh-modes cannot be driven so easily. An example is shown where a zigzag motion of a cosh-mode is induced. We have also used this control to induce collisions of two mutually displaced defect modes. The outcome depends on the defect trajectory after the collision being either a pair of driven defect modes or a single defect mode with stronger localization.

Finally we have discussed in detail how to excite different types of these defect modes in realistic scenarios.
We think that the mechanism proposed in this paper could provide one more possibility for experimentalists
 to manage gap solitons by using a small localized defect of an optical lattice.

\section*{Acknowledgments}
The authors are in debt to G.L. Alfimov who provided us with the  software which was used for obtaining some of the numerical results reported in this paper. 
V. A. B. acknowledges support of the FCT grant SFRH/BPD/5632/2001. V. A. B. and V. V. K. were supported by the FCT and FEDER under the grant POCI/FIS/56237/2004.  V.M.P-G. is partially supported by Ministerio de Educaci\'on y Ciencia under grant BFM2003-02832 and Consejer\'{\i}a de Educaci\'on y Ciencia de la Junta de Comunidades de Castilla-La Mancha under grant PAI-05-001.


\begin{thebibliography}{9}

\bibitem{Morsch} see e.g. O. Morsch, and M. Oberthaler, {\it Bose-Einstein condensates in optical lattices}, Rev. Mod. Phys. (to appear).

\bibitem{GapSol} B. Eiermann, Th. Anker, M.Albiez, M. Taglieber, P. Treutlein, K. P. Marzlin, and M. K. Oberthaler, Phys. Rev. Lett. {\bf 92}, 230401 (2004).


\bibitem{review} V. A. Brazhnyi and V. V. Konotop, Mod. Phys. Lett. B {\bf 18} 627 (2004). 


\bibitem{BKK}  V. A. Brazhnyi, V. V. Konotop, and V. Kuzmiak, Phys. Rev. A {\bf 70}, 043604   (2004).

\bibitem{10}C. Raman, M. K\"ohl, R. Onofrio, D. S. Durfee, C. E. Kuklewicz, Z. 
Hadzibabic and W. Ketterle, Phys. Rev. Lett. {\bf 83}, 2502 
(1999); 
R. Onofrio, C. Raman, J. M. Vogels, J. R. Abo-Shaeer, A. P. Chikkatur and W. Ketterle, Phys. Rev. Lett. 
\textbf{85}, 2228 (2000); C. Fort, L. Fallani, V. Guarrera, J. E. Lye, M. Modugno, D. S. Wiersma, M. Inguscio, Phys. Rev. Lett. \textbf{95}, 170410 (2005).

\bibitem{9}T. Winiecki, J. F. McCann and C. S. Adams, 
Phys. Rev. Lett. \textbf{82}, 5186 (1999).


\bibitem{11} V. Hakim, Phys. Rev. E \textbf{55}, 2835 (1997); 
A. Radouani, Phys. Rev. A  \textbf{70}, 013602 (2004).

\bibitem{AstPit} G. E. Astrakharchik, L.P. Pitaevskii, Phys. Rev. A \textbf{70}, 013608 (2004).

\bibitem{14} G. Herring, P.G. Kevrekidis, R. Carretero-Gonzalez, B.A. Malomed, D.J. Frantzeskakis, and
A. R. Bishop, Phys. Lett. A \textbf{345}, 144 (2005).

\bibitem{ours} V. A. Brazhnyi, V. V. Konotop, V. M. P\'erez-Garc\'{\i}a, Phys. Rev. Lett. {\bf 96}, 060403 (2006).

\bibitem{AKKS} G. L. Alfimov, P. G. Kevrekidis, V. V. Konotop, and M. Salerno,  Phys. Rev. E {\bf 66}, 046608 (2002). 

\bibitem{KAKS} V. V.  Konotop, G. L. Alfimov, P. G. Kevrekidis, and M. Salerno, 
 in ``Nonlinear Physics: Theory and Experiment. II'', Eds M. J. Ablowitz, M. Boiti, F. Pempinelli, and B. Prinari (World Scientific, 2003) p 280.

\bibitem{kohn} W. Kohn, Phys. Rev. {\bf 115}, 809 (1959). 

\bibitem{Goodman1} R. H. Goodman, R. E. Slusher, M. I. Weinstein, J. Opt. Soc. Am. B {\bf 19}, 1632 (2002).

\bibitem{Mak} W. C. K. Mak, B. A. Malomed, and P. L. Chu, J. Mod. Optics {\bf 51}, 2141 (2004).
 
\bibitem{dp1} P. G. Kevrekidis, Yu. S. Kivshar, and A. S. Kovalev, Phys. Rev. E {\bf 67}, 046604 (2003). 
 
\bibitem{Victor} V. M. P\'erez-Garc\'{\i}a, H. Michinel and H. Herrero, Phys. Rev. A {\bf 57}, 3837 (1998).
 
\bibitem{solid} See e.g., A. I. Anselm, {\it Introduction to semiconductor theory} (Moscow Englewood Cliffs, N.J., Prentice-Hall, 1981). 

\bibitem{KS} V. V. Konotop and M. Salerno, Phys. Rev. A {\bf 65}, 021602(R) (2002). 
 

 




\bibitem{Goodman}R. H. Goodman, P. J. Holmes, M. I. Weinstein,  Physica D \textbf{192}, 215 (2004).

\bibitem{classif} 
A.A. Sukhorukov, Yu. S. Kivshar, O. Bang, J. J. Rasmussen, and P. L. Christiansen, Phys. Rev. E {\bf 63}, 036601 (2001).

 

\bibitem{KPTV} V. V. Konotop, V. M. P\'erez-Garc\'{\i}a, Yi-Fa Tang, and L. V\'azquez, Phys. Lett. A {\bf 236}, 314 (1997). 

\bibitem{RSFS} S. Raghavan, A. Smerzi, S. Fantoni, and S. R. Shenoy, Phys. Rev. A {\bf 59}, 620 (1999).

\bibitem{PP} For a discussion about the link between the two methods see e. g. ~\cite{review}.

\end{thebibliography}
\end{document}